\documentclass[fleqn,usenatbib]{mnras}

\usepackage{newtxtext,newtxmath}

\usepackage[T1]{fontenc}
\usepackage{ae,aecompl}
\usepackage{natbib}
\usepackage{multicol}
\usepackage{xcolor}

\usepackage{graphicx}	
\usepackage{amsmath}	
\usepackage{enumerate}
\usepackage[caption=false]{subfig}

\usepackage{calc}
\usepackage{comment}
\usepackage{threeparttable}
\usepackage{xcolor}
\usepackage{lipsum}
\usepackage{amssymb}
\usepackage{macros}
\usepackage[shortlabels]{enumitem}
\usepackage{bm}

\newcommand{\rockstar}{\texttt{ROCKSTAR}}

\title[Dark-matter subhaloes near Milky Way-mass galaxies]{The dark side of FIRE: predicting the population of dark matter subhaloes around Milky Way-mass galaxies}

\author[Barry et al.]{Megan Barry$^{1}$\thanks{E-mail: \href{mailto:mlbarry@ucdavis.edu}{mlbarry@ucdavis.edu}},
Andrew Wetzel${^1}$,
Sierra Chapman$^{1}$, 
Jenna Samuel$^{2}$,
\newauthor
Robyn Sanderson$^{3,4}$,
Arpit Arora$^{3}$\\
$^{1}${Department of Physics \& Astronomy, University of California, Davis, 1 Shields Ave, Davis, CA 95616, USA} \\
$^{2}${Department of Astronomy, The University of Texas at Austin, 2515 Speedway, Stop C1400, Austin, TX 78712, USA}\\
$^{3}${Department of Physics and Astronomy, University of Pennsylvania, 209 South 33rd Street, Philadelphia, PA 19104, USA}\\
$^{4}${Center for Computational Astrophysics, Flatiron Institute, 162 Fifth Avenue, New York, NY 10010, USA}
}
\date{}

\pubyear{2023}

\begin{document}
\label{firstpage}
\pagerange{\pageref{firstpage}--\pageref{lastpage}}
\maketitle

\begin{abstract}
A variety of observational campaigns seek to test dark-matter models by measuring dark-matter subhaloes at low masses.
Despite their predicted lack of stars, these subhaloes may be detectable through gravitational lensing or via their gravitational perturbations on stellar streams.
To set measurable expectations for subhalo populations within $\Lambda$CDM, we examine 11 Milky Way (MW)-mass haloes from the FIRE-2 baryonic simulations, quantifying the counts and orbital fluxes for subhaloes with properties relevant to stellar stream interactions: masses down to $10^{6}\Msun$, distances $\lesssim50\kpc$ of the galactic center, across $z=0-1$ ($t_\textrm{lookback}=0-8\Gyr$).
We provide fits to our results and their dependence on subhalo mass, distance, and lookback time, for use in (semi)analytic models.
A typical MW-mass halo contains $\approx16$ subhaloes $>10^{7}\Msun$ ($\approx1$ subhalo $>10^{8}\Msun$) within 50 kpc at $z\approx0$.
We compare our results with dark-matter-only versions of the same simulations: because they lack a central galaxy potential, they overpredict subhalo counts by $2-10\times$, more so at smaller distances.
Subhalo counts around a given MW-mass galaxy declined over time, being $\approx10\times$ higher at $z=1$ than at $z\approx0$.
Subhaloes have nearly isotropic orbital velocity distributions at $z\approx0$. 
Across our simulations, we also identified 4 analogs of Large Magellanic Cloud satellite passages; these analogs enhance subhalo counts by $1.4-2.1$ times, significantly increasing the expected subhalo population around the MW today.
Our results imply an interaction rate of $\sim5$ per Gyr for a stream like GD-1, sufficient to make subhalo-stream interactions a promising method of measuring dark subhaloes.
\end{abstract}



\section{Introduction}
\label{sec:intro}

A key prediction of the \ac{CDM} model is the existence of effectively arbitrarily low-mass self-gravitating dark-matter (DM) structures, known as haloes, including subhaloes that reside within a more massive halo \citep{Klypin_1999, Moore_1999, Springel2008, Bullock2017}. Alternative models, such as warm dark matter (WDM) and fuzzy dark matter, predict a lower cutoff in the (sub)halo mass function \citep[for example][]{Hu_2000,  Ostdiek2022}. Current constraints on low-mass (sub)haloes come from luminous galaxies, such as the faint satellite galaxies around the \ac{MW}. Measurements of ultra-faint galaxies imply that (sub)haloes exist down to $\sim 10^{8} \Msun$ \citep[for example][]{Jethwa2018, Nadler2021}. Theoretical works show that (sub)haloes below this mass are below the atomic cooling limit and therefore unable to retain enough gas before cosmic reionization to support star formation, leaving them starless and thus invisible to direct detection \citep[for example][]{Bullock2000, Somerville_2002, Benson2002}. The discovery of completely dark (sub)haloes would represent another key success of the \ac{CDM} model, and such measurement (or lack thereof) would provide key constraints on the properties of dark matter.

To date, researchers have devised two potential avenues for indirectly detecting these dark (sub)haloes. One method uses gravitational lensing: the lensed light from a background galaxy allows us to determine a foreground galaxy's mass distribution \citep{Mao1998}, including low-mass (sub)haloes that reside along the line of sight. Most work using this method focuses on population statistics \citep{Sengul2022, Wagner2022, Ostdiek2022}, although  \citet{Vegetti2012, Vegetti2014} identified individual satellites with total mass $10^{8} - 10^{9} \Msun$ at $z \approx 0.2 - 0.5$. Existing works predominantly examine galaxies with DM halo masses $M_{\rm halo} \gtrsim 10^{13} \Msun$ at these redshifts, notably higher than \ac{MW}-mass galaxies with $M_{\rm halo} \approx 10^{12} \Msun$ at $z = 0$ \citep[for example][]{Bland_Hawthorn_2016}.

The \ac{MW} itself provides a separate means to measure dark subhaloes, via perturbations to thin streams of stars that originate from the tidal disruption of a globular cluster (GC) or satellite galaxy \citep{Ibata_2002, Johnston2002,Johnston_2016}. If a subhalo passes near such a stellar stream, its gravitational field can impart an identifiable gap, spur, or other perturbation, whose properties depend on the subhalo's mass, size, velocity, and other orbital parameters. Recent works explored how subhaloes with masses $\gtrsim 10^{5} \Msun$ can induce observable features in stellar streams \citep[for example][]{Yoon_2011, Carlberg2012, Erkal2016, Banik:2018pjp, Bonaca_2019}; less massive subhaloes lack the energy necessary to leave observable evidence of interaction. To confirm that a dark subhalo induced a particular perturbation, one must rule out the effects of luminous objects, including the MW's $> 50$ known satellite galaxies \citep{McConnachie_2012, Simon2019} and $> 150$ known GCs \citep{Harris2010}, as well as giant molecular clouds \citep{Amorisco_2016} and other stellar streams \citep{Dillamore_2022}.

Of the dozens of currently well-known streams around the \ac{MW} \citep[for example][]{Grillmair_2016,Mateu23}, most studies focus on two, GD-1 and Pal 5, given their relative proximity to the \ac{MW} and the high-quality 6D phase-space data available for them.
GD-1 is $\approx 15 \kpc$ long, with a pericenter of 13 kpc and apocenter of 27 kpc, and it formed $\approx 3 \Gyr$ ago \citep{Doke2022}, likely from a progenitor GC \citep{Bonaca2018}. Pal 5 is $\approx 10 \kpc$ long \citep{Starkman_2020}, with a pericenter of 8 kpc and an apocenter of 19 kpc \citep{Yoon_2011}, and it formed $\approx 8 \Gyr$ ago from the Pal 5 GC \citep{Odenkirchen2001}.
GD-1 and Pal 5 represent perhaps the ideal streams on which to study the potential gravitational impacts of dark subhaloes, though \textit{Gaia} data release 3 (DR3) now provides even more detailed 6D phase-space measurements of stars in more streams \citep{Gaia}.

In addition to the subhaloes orbiting the \ac{MW} itself, the \ac{CDM} model also predicts that large satellites such as the \ac{LMC} host their own orbiting subhaloes \citep[for example][]{Deason2015, Sales_2016, Jahn2019, Santos2021}. Given that the \ac{LMC} just passed its pericenter of $d \approx 50 \kpc$ from the \ac{MW} center \citep{Kallivayalil2013}, the inner halo currently may be in a temporary period of enhanced subhalo enrichment \citep{Dooley2017b, Dooley2017}.

Theoretical predictions for the counts, orbits, and sizes of dark subhaloes can help support or rule out a dark subhalo origin for observed gaps or other features in stellar streams.
Previous works have predicted subhalo populations in the mass regime relevant to subhalo-stream interactions ($\approx 10^{5} - 10^{8} \Msun$). Most used dark-matter-only (DMO) simulations that do not account for the effects of baryonic matter \citep[for example][]{Yoon_2011, Mao2015, Griffen2016}. However, incorporating baryonic physics significantly can affect subhalo populations. Primarily, the presence of the central galaxy induces additional tidal stripping on subhaloes that orbit near it, which, as previous works \citep[for example][]{DOnghia10, GK2017, Webb_2020} showed, causes DMO simulations to overpredict subhalo count significantly, by $5 \times$ or more, near the central galaxy. Additionally, gas heating from the cosmic UV background reduces the initial masses and subsequent accretion history of low-mass (sub)haloes, making them lower mass today \citep[for example][]{Bullock2000, Somerville_2002, Benson2002}.

The FIRE-2 cosmological zoom-in simulations are well suited for predicting the population of low-mass dark subhaloes, given their high resolution and inclusion of relevant baryonic physics; most importantly, the formation of realistic \ac{MW}-mass galaxies.
As a critical benchmark, previous works have shown that the luminous subhaloes (satellite galaxies) around \ac{MW}-mass galaxies in FIRE-2 broadly match the distributions of stellar masses and internal velocities \citep{Wetzel2016, GK2019}, as well as radial distance distributions \citep{Samuel2020}, of satellite galaxies observed around the \ac{MW} and M31, as well as \ac{MW} analogs in the SAGA survey \citep{Geha_2017}. Furthermore, as \citet{Samuel_2021} showed, FIRE-2 \ac{MW}-mass galaxies with an \ac{LMC}-like satellite show much better agreement with various metrics of satellite planarity, as observed around the \ac{MW} and M31, which motivates further exploration of potential effects of the \ac{LMC} on the population of low-mass, dark subhaloes.

In this work, we extend these FIRE-2 predictions of satellite populations down to lower-mass subhaloes, with DM masses as low as $10^{6} \Msun$, which, as we described above, are likely to be completely dark (devoid of stars).
We examine subhaloes within 50 kpc of \ac{MW}-mass galaxies, the regime most relevant for observable interactions of dark subhaloes with stellar streams. We expand in particular on the work of \cite{GK2017}; we examine subhaloes across a much larger set of 11 \ac{MW}-mass haloes (instead of 2), and we time-average over multiple simulation snapshots (instead of just the one at $z = 0$) for improved statistics for the small number of subhaloes that survive near the \ac{MW}-mass galaxy.

\section{Methods}
\label{sec:sim}

\subsection{FIRE-2 simulations of \ac{MW}-mass haloes}
\label{sec:fire2}

\begin{table*}
\centering
\caption{
Properties of the \ac{MW}/M31-mass galaxies/haloes at $z \approx 0$ in the FIRE-2 simulations. We include galaxies with $\Mstar = 2.5 - 10 \times 10^{10} \Msun$, within a factor of $\approx 2$ of the \ac{MW}. The last 3 columns list the number of subhaloes above the given threshold in instantaneous dark-matter mass that are within 50 kpc of the host, time-averaged across $z = 0 - 0.15$ (1.9 Gyr). 
[1]: \protect\cite{Hopkins2018}, 
[2]: \protect\cite{GK2019}, 
[3]: \protect\cite{GK2017}, 
[4]: \protect\cite{Wetzel2016}, 
[5]: \protect\cite{Samuel2020}
}
\begin{tabular}{llllllll}
\hline
\multicolumn{1}{|c|}{Name} & 
\multicolumn{1}{|c|}{$\Mstar$}  & 
\multicolumn{1}{|c|}{$M_{200\rm{m}}$}  & 
\multicolumn{1}{|c|}{$M_{\rm total}(< 100 \kpc)$}  & 
\multicolumn{1}{|c|}{$N_{\rm{subhalo}}$} & 
\multicolumn{1}{|c|}{$N_{\rm{subhalo}}$} & 
\multicolumn{1}{|c|}{$N_{\rm{subhalo}}$} &
\multicolumn{1}{|c|}{Introduced in} \

\cr &

[$10^{10} \Msun$] &
[$10^{12} \Msun$] &
[$10^{11} \Msun$] & 
$(>10^6 \Msun)$ &
$(>10^7 \Msun)$ &
$(>10^8 \Msun)$
\\
\hline
m12m    & 10.0 & 1.6 & 7.9 & 123 & 21.3 & 1.4 & [1]\\
Romulus &  8.0 & 2.1 & 10.4 & 143 & 16.4 & 1.8 & [2]\\
m12b    &  7.3 & 1.4 & 7.2 &  90 & 14.0 & 0.9 & [2]\\
m12f    &  6.9 & 1.7 & 8.0 & 106  & 18.2 & 1.4 & [3] \\
Thelma  &  6.3 & 1.4 & 7.7 & 179 & 30.9 & 2.7 & [2]\\
Romeo   &  5.9 & 1.3 & 7.7 & 168 & 18.9 & 1.6 & [2]\\
m12i    &  5.5 & 1.2 & 6.4 & 131 & 20.4 & 1.9 & [4]\\
m12c    &  5.1 & 1.4 & 6.2 & 246 & 47.4 & 4.2 & [2]\\
m12w    &  4.8 & 1.1 & 5.5 & 162 & 18.8 & 1.9 & [5]\\
Remus   &  4.0 & 1.2 & 6.8 & 132 & 20.4 & 2.3 & [2]\\
Juliet  &  3.4 & 1.1 & 5.9 & 207 & 29.2 & 2.5 & [2]\\
\hline
average & 6.1 & 1.4 & 5.9 & 153 & 23.3 & 2.1
\end{tabular}
\label{hosttable}
\end{table*}

We analyze simulated host galaxy haloes from the FIRE-2 cosmological zoom-in simulations \citep{Hopkins2018}. We generated each simulation using \textsc{Gizmo} \citep{Hopkins2015}, which models $N$-body gravitational dynamics with an updated version of the \textsc{GADGET-3} TreePM solver \citep{Springel2005}, and hydrodynamics via the meshless finite-mass method. FIRE-2 incorporates a variety of gas heating and cooling processes, including free-free radiation, photoionization and recombination, Compton, photo-electric and dust collisional, cosmic ray, molecular, metal-line, and fine-structure processes, including 11 elements (H, He, C, N, O, Ne, Mg, Si, S, Ca, Fe). We use the model from \cite{Faucher2009} for the cosmic UV background, in which HI reionization occurs at $z \approx 10$. Each simulation consists of dark-matter, star, and gas particles. Star formation occurs in gas that is self-gravitating, Jeans-unstable, cold ($T < 10^{4}$ K), dense ($n > 1000 \cci$), and molecular as in \cite{krumholz}. Once formed, star particles undergo several feedback processes, including core-collapse and Type Ia supernovae, continuous stellar mass loss, photoionization and photoelectric heating, and radiation pressure.

We generated cosmological initial conditions for each simulation at $z \approx 99$, within periodic boxes of length $70.4 - 172 \Mpc$ using the MUSIC code \citep{MUSIC}. We assume flat $\Lambda$CDM cosmologies, with slightly different parameters across our host selection: $h = 0.68 - 0.71$, $\Omega_{\rm \Lambda} = 0.69 - 0.734$, $\Omega_{\rm m} = 0.266-0.31$, $\Omega_{\rm b} = 0.0455 - 0.048$, $\sigma_{8} = 0.801 - 0.82$, and $n_{\rm s} = 0.961 - 0.97$, broadly consistent with \cite{Planck}. We saved 600 snapshots from $z=99$ to $0$, with typical spacing $\lesssim 25 \Myr$.

We examine host haloes from two suites of simulations. The first is the \textit{Latte} suite of individual MW-mass haloes \citep[introduced in][]{Wetzel2016}, which have dark-matter halo masses of $M_{200\textrm{m}} = 1 - 2 \times 10^{12} \Msun$ and no neighboring haloes of similar or greater mass within at least $\approx 5 \Rthm$, where $\Rthm$ is defined as the radius within which the density is 200 times the mean matter density of the Universe. Gas cells and star particles have initial masses of $7070 \Msun$, while dark-matter particles have a mass of $3.5 \times 10^{4} \Msun$. \textit{Latte} uses gravitational force softening lengths of 40 pc for dark matter and 4 pc for star particles (comoving at $z > 9$ and physical thereafter), and the gravitational softening for gas is adaptive, matching the hydrodynamic smoothing, down to 1 pc.
We also examine host haloes from the ELVIS on FIRE suite of Local Group-like MW+M31 halo pairs \citep[introduced in][]{GK2019}. These simulations have $\approx 2 \times$ better mass resolution than \textit{Latte}: Romeo \& Juliet have initial masses of $3500 \Msun$ for baryons and $1.9 \times 10^{4} \Msun$ for dark matter, while Romulus \& Remus and Thelma \& Louise have initial masses of $4000 \Msun$ for baryons and $2.0 \times 10^{4} \Msun$ for dark matter. ELVIS uses gravitational force softening lengths of $\approx 32$ pc for dark matter, $2.7 - 4.4$ pc for stars, and$ 0.4 - 0.7$ pc (minimum) for gas.

To ensure similarity to the \ac{MW}, we selected host galaxies from these suites that have a stellar mass within a factor of $\approx 2$ of  $M_{\textrm{MW}} \approx 5 \times 10^{10} \Msun$ \citep{Bland_Hawthorn_2016}, which leaves 11 total hosts: 6 from \textit{Latte} and 5 from ELVIS. Table~\ref{hosttable} lists their properties at $z \approx 0$. For each simulation, we also generated a DMO version at the same resolution, and we compare these against our baryonic simulations to understand the effects of baryons on subhalo populations.
The primary effect of baryons for our analysis of low-mass subhaloes is simply the additional gravitational potential of the \ac{MW}-mass galaxy \citep{GK2017}.

\subsection{Finding and measuring subhaloes}
\label{sh_select}

We examine subhaloes, which we define as lower-mass haloes that reside within a $\Rthm$ of a \ac{MW}-mass host halo. We identify dark-matter subhaloes using the \textsc{Rockstar} 6D halo finder \citep{Behroozi2012a}, defining (sub)haloes as regions of space with a dark matter density $>200\times$ the mean matter density. We include subhaloes that have a bound mass fraction of $> 0.4$ and at least 30 dark-matter particles, then construct merger trees using \textsc{CONSISTENT-TREES} \citep{Behroozi2012b}. For numerical stability, we first generate (sub)halo catalogs using only dark-matter particles, then we assign star particles to haloes in post-processing \cite[see][]{Samuel2020}.

From the merger trees, we select subhaloes with masses, distances, and redshifts that are most relevant for observable gravitational interactions with stellar streams \citep[as in][]{Koposov2010, Thomas_2016, Li2022}.
Throughout, we examine subhaloes according to their \textit{instantaneous} mass, given that this mass, rather than the pre-infall `peak' mass, is more relevant for the strength of stream interactions (or the strength of gravitational lensing perturbations).
We examine three thresholds in instantaneous mass: $M_{\rm sub} >10^{6}$, $> 10^{7}$, and $> 10^{8} \Msun$, corresponding to a minimum of $\approx 30 - 60$, $300 - 600$, and $3000 - 6000$ dark-matter particles, being lower in the Latte simulations and higher in the ELVIS Local Group-like simulations. These subhaloes typically had $\gtrsim 4 \times$ higher mass (more dark-matter particles) prior to \ac{MW} infall and tidal mass stripping, independent of subhalo mass. When examining subhaloes in DMO simulations, we reduce their masses by the cosmic baryon mass fraction ($\approx 15$ per cent), assuming that these subhaloes would have lost essentially all of their baryonic mass, consistent with the properties of subhaloes at these masses in our baryonic simulations \citep[see also][]{Bullock2017}.

We include all subhaloes above these mass thresholds regardless of whether they are luminous or dark.
For subhaloes within $50 \kpc$, the fraction that contain at least 6 star particles (the limit of our galaxy catalog) is: 30 per cent at $M_{\rm sub} > 10^{8} \Msun$, 5 per cent at $M_{\rm sub} > 10^{7} \Msun$, and $\lesssim 1$ per cent at $M_{\rm sub} > 10^{6} \Msun$.

In Appendix~\ref{sec:app}, we explore the resolution convergence of our results. In summary, our tests show that the counts of subhaloes with $M_{\rm sub} > 10^{7} \Msun$ are well converged, but our simulations likely underestimate subhalo counts at $M_{\rm sub} > 10^{6} \Msun$ by up to $\approx 1.5 - 2$ (depending on distance) at $z \approx 0$, so one should consider those results as lower limits to the true counts.
We show results for $M_{\textrm{sub}} > 10^{6} \Msun$ in a lighter shade to reinforce this caution.

We use three metrics to quantify subhalo counts: number enclosed, number density, and orbital radial flux.
The number enclosed, $N(< d)$, includes all subhaloes within a given distance $d$ from the host galaxy center.
We calculate the subhalo number density, $n(d)$, by counting all subhaloes within a spherical shell 5 kpc thick, with the shell midpoint centered at $d$, and dividing by the volume of the shell.
To inform predictions for subhalo-stream interaction rates, we also include the subhalo orbital radial flux, $f(d)$, or the amount of subhalos passing into or out of a host-centered spherical surface of radius $d$ per Gyr. We count a subhalo as passing through the surface if, between two adjacent snapshots, it orbited from outside to inside the surface or vice versa. We do not distinguish between inward and outward flux. 
While our snapshot spacing of $20 - 25 \Myr$ provides good time resolution, we also interpolate the distances of subhaloes between snapshots. For each subhalo within 5 kpc of a given distance bin, we apply a cubic spline fit to its distance from the host for several snapshots surrounding the current snapshot to determine its distance between snapshots. Using these interpolated distances is important at small $d$, where the surface-crossing times are shortest: it increases the measured flux by $\approx 20$ per cent at $d < 10 \kpc$ compared to using the snapshots alone.
We will also show that, at $z\approx 0$, subhaloes have approximately isotropic orbits on average, so our values for radial flux can be used as flux rates in any direction at low redshifts.

We examine trends back to $z = 1$ (lookback time $t^\textrm{lb} \approx 8 \Gyr$), because observable dynamical perturbations to stellar streams could have occurred several Gyr ago \citep[see][]{Yoon_2011}. Furthermore, because subhalo counts are subject to time variability and Poisson noise, especially at small distances, given that an orbit spends the least time near pericenter, we follow the approach in \cite{Samuel2020}: for each host halo, we time-average its subhalo population across 92 snapshots at $z = 0 - 0.15$ ($t^\textrm{lb} = 0 - 2.15 \Gyr$).
We then compute the median and 68 per cent distribution across the 11 host haloes. When examining redshift evolution, we average over 3 redshift ranges: $z = 0.0 - 0.1$ ($t^\textrm{lb} = 0 - 1.3$ Gyr, 66 snapshots), $z = 0.5 - 0.6$ ($t^\textrm{lb} = 5.1 - 5.7 \Gyr$, 25 snapshots), and $z = 1.0 - 1.1$ ($t^\textrm{lb} = 7.8 - 8.2 \Gyr$, 14 snapshots).

We use the publicly available Python packages, \textsc{GizmoAnalysis} \citep{Wetzel_Gizmo} and \textsc{HaloAnalysis} \citep{Wetzel_Halo} to analyze these data.

\subsection{LMC satellite analogs}
\label{sec:lmcselection}

\begin{table*}
\centering
\caption{
Properties of the four \ac{LMC} satellite analogs, at the lookback time that each one first orbits within a distance of $50 \kpc$ from their \ac{MW}-mass host galaxy.
See Section~\ref{sec:lmcselection} for details on their selection criteria.
While we list the actual (initial) pericentric distance of their orbit, we present all results when these satellites first were at $d \approx 50 \kpc$, to provide context for the \ac{LMC} at its current distance.
$M_{\rm sub,inst}$ indicates the instantaneous subhalo dark-matter mass of the \ac{LMC} analog this time, while $M_\textrm{sub,peak}$ is its peak (sub)halo mass throughout its history.
}
\begin{tabular}{lllllll}
\hline
Host &
$t^{\textrm{lb}}_{\rm 50 \ kpc}$ [Gyr] &
$z_{\rm 50 \ kpc}$ &
$\Mstar$ [$10^{9} \Msun$] &
$M_{\rm sub,inst}$ [$10^{11} \Msun$] &
$M_{\rm sub,peak}$ [$10^{11} \Msun$] &
$d_{\rm peri}$ [kpc] \\
\hline
m12w & 6.0 & 0.60 & 1.25 & 0.9 & 1.3 &  8 \\
m12b & 5.1 & 0.50 & 7.13 & 1.7 & 2.1 & 38 \\
m12f & 3.1 & 0.27 & 2.62 & 1.1 & 1.6 & 36 \\
m12c & 1.0 & 0.08 & 1.17 & 1.2 & 1.7 & 18 \\
\hline
\end{tabular}
\label{tab:LMC-table}
\end{table*}

Numerous works have demonstrated the likely contribution of the \ac{LMC} to the population of luminous satellite galaxies around the \ac{MW} \citep[for example][]{Hargis2014, Deason2015, Sales_2016, Jethwa2016, Dooley2017b, Dooley2017, Patel_2020}.
This motivates the possibility that the \ac{LMC} also may have contributed a significant fraction of non-luminous lower-mass subhaloes as well. To assess if the presence of the LMC today affects predictions for subhaloes close to the MW, we select host haloes that contain a satellite that is an analog to the LMC, following \cite{Samuel_2022}, with the following constraints:

\begin{enumerate}[leftmargin=*]
\item Pericentric passage at $t^\mathrm{lb} < 6.4 \Gyr$ ($z < 0.7$): we choose this broad time window to capture a larger number of (rare) \ac{LMC}-like passages.
\item $M_\mathrm{sub,peak} > 4 \times 10^{10} \Msun$ or $\Mstar > 5 \times 10^{8} \Msun$: consistent with observations and inferences of the \ac{LMC}'s mass \citep[see][]{Erkal2019,Vasiliev2021}.
\item $d_\mathrm{peri} < 50 \kpc$: consistent with the current measured pericentric distance of the \ac{LMC} \citep[see][]{Kallivayalil2013}.
\item The satellite is at its first pericentric passage, consistent with several lines of evidence that suggest that the LMC is on its first infall into the MW \citep[see][]{Kallivayalil2013, Sales_2016}.
\end{enumerate}

From our 11 MW-mass haloes, this leaves 4 LMC analogs that meet all four criteria. Table~\ref{tab:LMC-table} lists their properties, including masses and pericenters. Because we are interested in how the LMC affects recent MW subhalo populations, we show properties of each LMC satellite analog when it first reached a distance of 50 kpc from the galaxy center, corresponding to the LMC's current distance.

\section{Results}
\label{sec:res}

\subsection{Counts and orbital radial fluxes}

\begin{figure}
\centering
\begin{tabular}{c}
\includegraphics[width = 0.94 \linewidth]{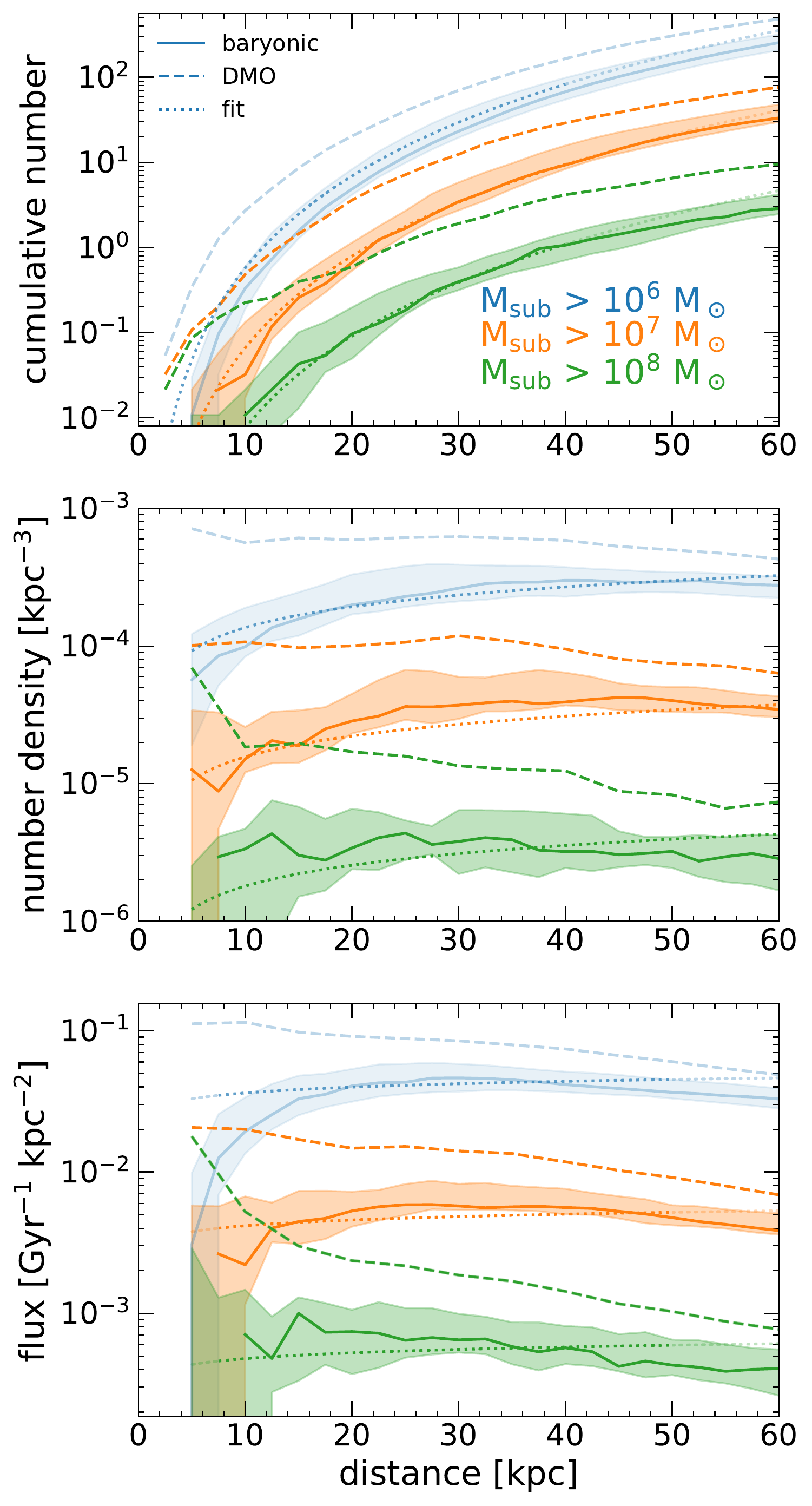}
\end{tabular}
\vspace{-3 mm}
\caption{
Counts and orbital radial fluxes of dark-matter subhaloes as a function of distance, $d$, from the central \ac{MW}-mass galaxy at $z \approx 0$.
For each halo, we time-average its subhalo population over 92 snapshots across $z = 0 - 0.15$ (1.9 Gyr).
Solid lines and shaded regions show the median and $68$ per cent distribution across the 11 host haloes.
We show results for $M_{\rm sub} > 10^{6} \Msun$ in a lighter shade to indicate potential resolution effects (see Section~\ref{sh_select}).
Dashed lines show dark-matter-only (DMO) simulations of the same haloes.
Dotted lines show the fits in Table~\ref{tab:fit-parameters}; a lighter shade indicates points outside of the distance range used for fitting.
\textbf{Top}: Cumulative number of subhaloes, $N(< d)$, within sphere of radius $d$.
\textbf{Middle}: Number density, $n(d)$, of subhaloes within a spherical shell $\pm 2.5$ kpc of $d$.
\textbf{Bottom}: Orbital radial flux of subhaloes, that is, the number of subhaloes per kpc$^2$ per Gyr passing either inwards or outwards through a spherical surface of radius $d$.
Because of the additional gravitational tidal stripping from the \ac{MW}-mass galaxy in the baryonic simulations (unlike in the DMO simulations), both $n(d)$ and flux vary only weakly with $d$, to within a factor of a few.}
\label{masspop}
\end{figure}

\begin{figure}
\centering
\begin{tabular}{c}   
\includegraphics[width = 0.94 \linewidth]{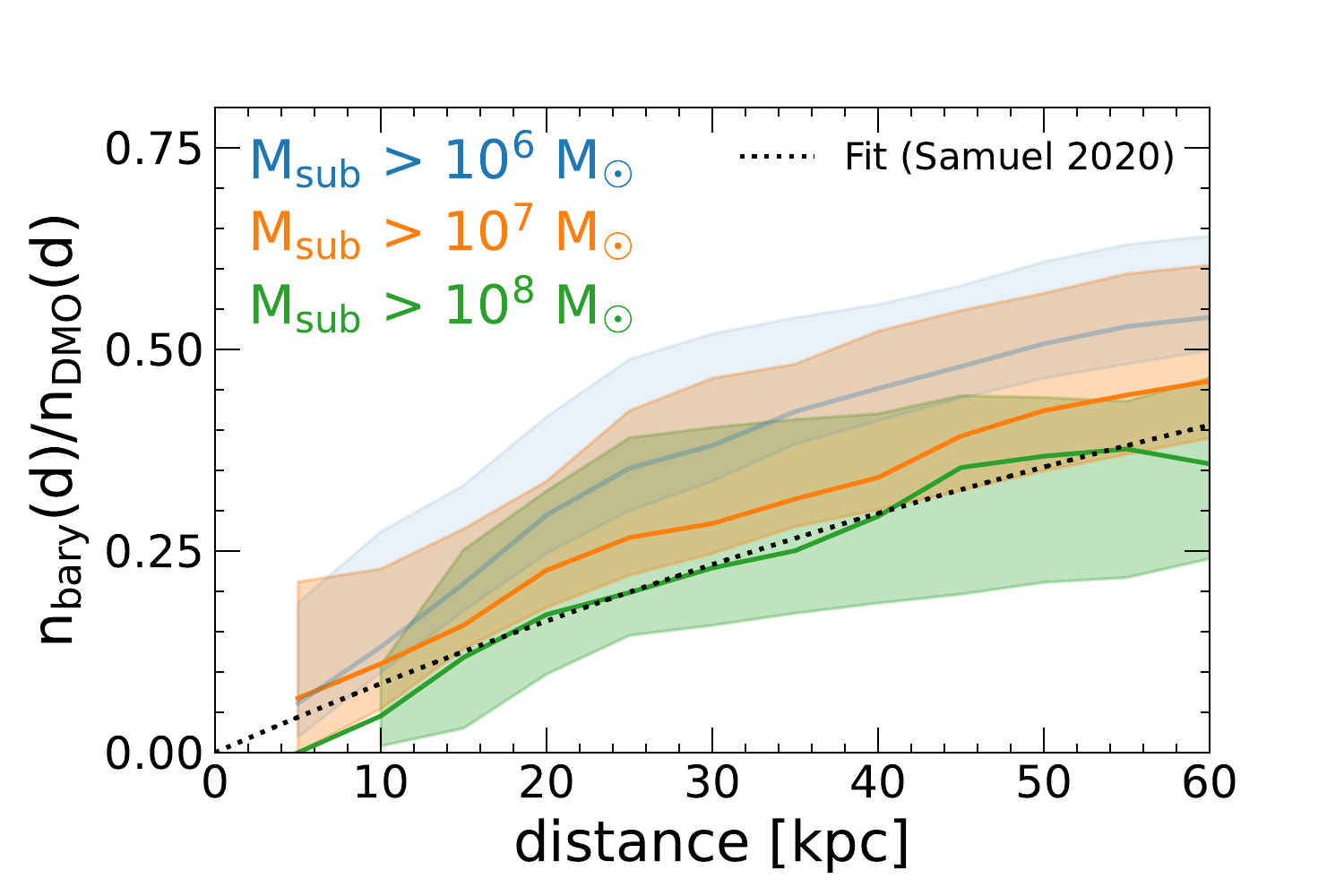}
\end{tabular}
\vspace{-2 mm}
\caption{
Ratio of the number density of subhaloes at $z \approx 0$ (as in Figure~\ref{masspop}, middle) in baryonic simulations relative to dark-matter-only (DMO) versions of the same host haloes.
We show the median and $68$ per cent distribution across the 11 haloes.
This ratio illustrates the significant depletion of subhaloes in baryonic simulations, especially at small $d$, primarily from the increased gravitational tidal stripping from the presence of the \ac{MW}-mass galaxy. The typical ratio at $d = 10 - 30 \kpc$, the approximate distances of the GD-1 and Pal-5 streams, is $0.05 - 0.25$, so DMO simulations overpredict subhalo counts by $4 - 20 \times$.
We show the fit from \protect\cite{Samuel2020} for more massive (luminous) subhaloes, with $M_\textrm{sub,peak} > 8 \times 10^{8} \Msun$, as a dotted line, which matches well our results at $M_\textrm{sub} > 10^{8} \Msun$.
}
\label{masspop_ratio}
\end{figure}

In Figure~\ref{masspop}, we characterize subhalo counts via three metrics: (1) the cumulative number of subhaloes, $N(< d)$, within a spherical shell at distance $d$ from the \ac{MW}-mass galaxy center; (2) the number density of subhaloes, $n(d)$, within $\pm 2.5$ kpc of $d$, and (3) the orbital radial flux of subhaloes through a spherical surface at $d$, including both incoming and outgoing subhaloes.
We show each  metric for three thresholds in subhalo instantaneous dark-matter mass: $M_{\textrm{sub}} > 10^{6}$, $10^{7}$, and $10^{8} \Msun$.
Here and throughout, we show results for $M_\textrm{sub} > 10^{6} \Msun$ with a lighter shade, to emphasize that those counts are likely lower limits, given the resolution considerations in Appendix~\ref{sec:app}.

All three metrics of subhalo counts decrease roughly linearly with increasing subhalo mass at a given $d$.
Within the approximate orbital distances of GD-1 and Pal 5, $d \approx 10 - 30 \kpc$ \citep{Price_Whelan_2018}, we predict $\approx 4$ subhaloes of $M_\textrm{sub} > 10^{7} \Msun$ and at least 20 subhaloes of $M_\textrm{sub} > 10^{6} \Msun$. We find no significant differences in subhalo counts between \textit{Latte} hosts and ELVIS hosts. Interestingly, both number density and flux vary only weakly with $d$, to within a factor of a few.
This is in contrast with the DMO simulations, which show a strong rise in these quantities towards smaller $d$.
Unlike \cite{GK2017}, who analyzed only the snapshot at $z=0$, our averaging across multiple snapshots reveals significant populations of subhaloes at small $d$.

Figure~\ref{masspop_ratio} quantifies the differences between the baryonic and DMO simulations, showing the ratio of the number density, $n(d)$, at a given $d$, given that many previous works used DMO simulations to explore subhalo populations.
Subhalo counts in baryonic simulations are systematically smaller than those in DMO, especially at small $d$, primarily because of the additional gravitational tidal stripping from the \ac{MW}-mass galaxy \citep[for example][]{GK2017, Kelley_2019}. DMO simulations overpredict subhalo counts by $\approx 2 - 3 \times$ at $d \approx 50 \kpc$ and by an order of magnitude at $d \lesssim 10 \kpc$, across all mass thresholds. At our fiducial stellar stream distances of $d \approx 10 - 30 \kpc$, the ratio is $\approx 0.05 - 0.25$ (that is, $4 - 20 \times$ more subhaloes in DMO simulations). These results are similar to \cite{Samuel2020}, who analyzed more-massive, luminous subhaloes with $M_\textrm{peak} > 8 \times 10^{8} \Msun$ in the same simulations, and found good agreement in the radial distance distribution with observations of satellites around the \ac{MW} and M31. Figure~\ref{masspop_ratio} shows their fit for the ratio of luminous satellites to DMO subhaloes via the dotted line, being $\approx 0.3$ at $d \approx 50 \kpc$.

\begin{figure}
\centering
\begin{tabular}{c}
\includegraphics[width = 0.94 \linewidth]{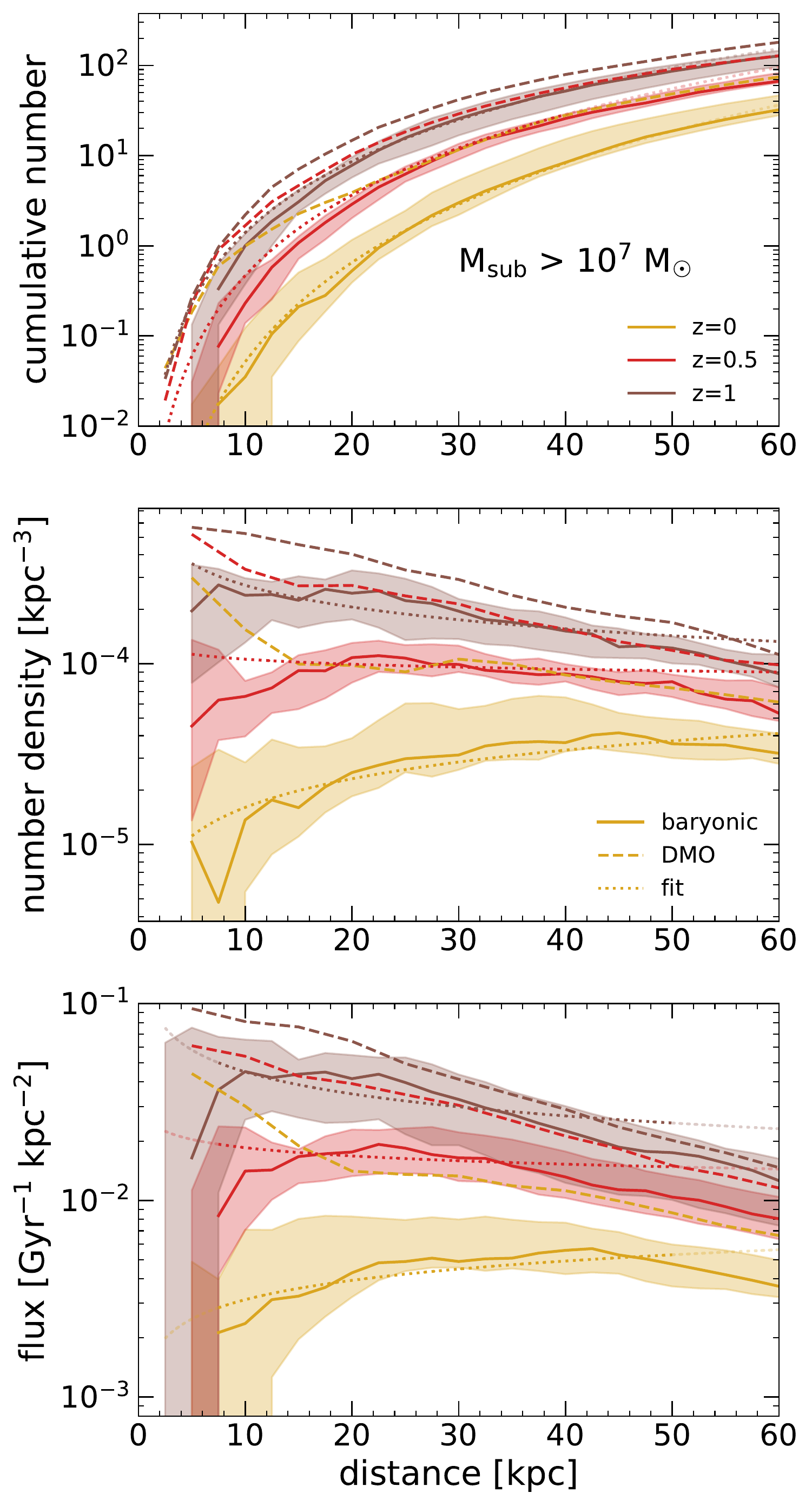}
\end{tabular}
\vspace{-3 mm}
\caption{
Counts and orbital radial flux versus distance, $d$, from the \ac{MW}-mass galaxy, for dark-matter subhaloes with $M_{\rm sub} > 10^{7} \Msun$ at different redshifts. We show the median and $68$ per cent distribution across the 11 host haloes. We time-average each one over the range $z = 0 - 0.1$, $0.5 - 0.6$, and $1 - 1.1$, corresponding to lookback times $0 - 1.3$, $5.1 - 5.7$, and $7.8 - 8.2 \Gyr$.
Dashed lines show median values for dark-matter-only (DMO) simulations of the \textit{Latte} hosts.
Dotted lines show the fits from Table~\ref{tab:fit-parameters}; a lighter shade indicates points outside of the distance range used for fitting.
\textbf{Top}: Cumulative number, $N(< d)$, within a sphere of radius $d$.
\textbf{Middle}: Number density, $n(d)$, within a spherical shell $\pm 2.5$ kpc of $d$.
\textbf{Bottom}: Orbital radial flux, that is, the number of subhaloes per Gyr passing in and out of a spherical surface of radius $d$.
All subhalo counts decrease over cosmic time, by up to $\approx 20 \times$ at $d = 10 \kpc$, and less dramatically ($\sim 3 \times$) at $d = 50 \kpc$.
The counts of subhaloes in baryonic simulations decreased more dramatically than in DMO simulations ($\approx 4 \times$ at most), especially at small $d$, because of additional tidal stripping from the \ac{MW}-mass galaxy.
}
\label{zpop}
\end{figure}

Figure~\ref{zpop} shows the same metrics as Figure~\ref{masspop}, for only subhaloes with $M_{\rm sub} > 10^{7} \Msun$, at 3 redshifts: $z \approx 0$, $\approx 0.5$, and $\approx 1$ (still averaged across multiple snapshots, see Section~\ref{sh_select}). All subhalo counts decreased over cosmic time at a given $d$; from $z = 1$ to $z = 0$, subhalo number density declined by a factor of $\approx 15$. We expect such a decline because, as explored in \cite[for example][]{Wetzel2009}, subhalo merging and destruction rates at earlier times are faster than the infall rate of new subhaloes, given the decline in overall accretion rate as the Universe expands. This decline occurred in both baryonic and DMO simulations, but at small $d$, the decrease over time is more significant in the baryonic simulations, given the higher rates of tidal stripping from the \ac{MW}-mass galaxy. We find decreases of $\approx 20 \times$ at $d = 10$ kpc and $\approx 3 \times$ at $d = 50$ kpc from $z=1$ to $z=0$.

\begin{figure}
\centering
\begin{tabular}{c}
\includegraphics[width = 0.94 \linewidth]{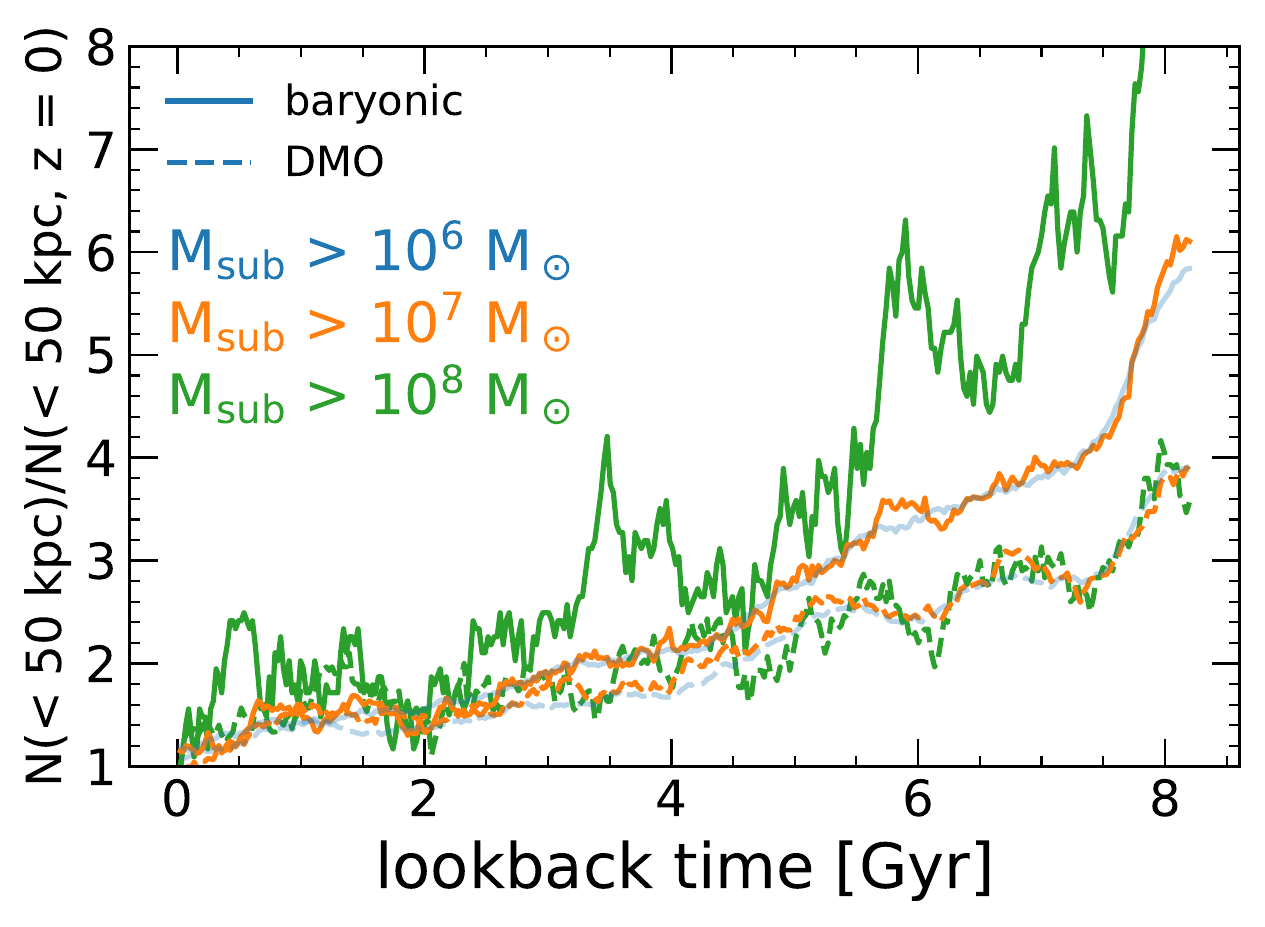}
\end{tabular}
\vspace{-3 mm}
\caption{
Ratio of the cumulative number of subhaloes enclosed within 50 kpc at varying redshifts to the number at $z=0$, as a measure of the relative depletion of subhaloes over cosmic time.
Solid lines show the mean across the 11 host haloes, while dashed lines show dark-matter-only (DMO) simulations of the same haloes. Dashed lines for $t^{\rm lb}> 5 \Gyr$ show only \textit{Latte} hosts.
We show results for $M_{\rm sub} > 10^{6} \Msun$ in a lighter shade to indicate potential resolution effects (see Section~\ref{sh_select}).
Since $z = 1$ ($t^\textrm{lb} \approx 8 \Gyr$), the subhalo population in baryonic simulations has decreased by $5 - 10 \times$, especially at higher subhalo masses.
Subhalo counts at $M_\textrm{sub} > 10^{8}$ were as high as $10 \times$ their values today, at $t^\textrm{lb} \gtrsim 8 \Gyr$ (extending above the axis).
}
\label{z_ratio}
\end{figure}

Figure~\ref{z_ratio} quantifies this decrease over time, for the three mass thresholds, via the ratio of $N(< 50 \kpc)$ at a given lookback time to the same value today. In baryonic simulations, this decrease has a slight mass dependence, while declines are similar across all masses in the DMO versions. Thus, using subhalo counts only at $z = 0$ would underestimate the subhalo population averaged across the last several Gyr, especially if the observable impacts on stellar streams persist for several Gyr. Our results at higher redshift also inform typical gravitational lensing studies, given that most observed lenses for measuring (sub)halo populations are at $z > 0$.

\begin{table}
\centering
\caption{
Fit parameters to Equation~\ref{eq:fit} for subhalo counts and orbital radial fluxes, where $a$ is the expansion scale factor, $d$ is the distance from the MW-mass galaxy in kpc, with $d_{0} = 1 \kpc$ as unit normalization, and $M$ is the lower limit on subhalo instantaneous dark-matter mass in $\Msun$, with $M_{0} = 10^{7} \Msun$ as unit normalization. Cumulative number, $N(< d)$ represents the total number of subhaloes enclosed in a sphere of radius $d$ centered on the \ac{MW}-mass galaxy; number density, $n(d)$, represents the subhalo density in a spherical shell within $\pm 2.5$ kpc of $d$; and radial flux, $f(d)$ represents the number of subhaloes per area that cross into or out of $d$ per Gyr. Accounting for the presence of the LMC (see Section~\ref{sec:LMC}) boosts these counts by $\approx 1.4 - 2.1 \times$.
}
\begin{tabular}{llllll} 
\hline
$p(>M, a)$ & $c_0$ & $c_1$ & $c_2$ & $c_3$ & $c_4$ \\
\hline
N(< d) & $1.24 \pm 0.62$ & 12.10 & 2.21 & 1.54 & -0.94 \\
n(d) [kpc$^{-3}$] & $0.12\pm 0.04$ & 10.53 & 1.97 & -1.36 & -0.94 \\
f(d) [Gyr$^{-1}$ kpc$^{-2}$] & $8.94\pm 2.82$ & 9.08 & 1.48 & -1.10 & -0.94 \\
\hline
\end{tabular}
\label{tab:fit-parameters}
\end{table}

For use in (semi)analytic models, we fit the three metrics---cumulative number, $N(< d)$, number density, $n(d)$, and orbital flux, $f(d)$---to the functional form
\begin{equation}
\; \; \; \; \; \; \; \; \; \; \; \; \; \; \; \; p(>M, d, a) = c_0 e^{- c_1 a}  \Big (\frac{d}{d_{0}}\Big )^{c_2 a + c_3}  \Big (\frac{M}{M_{0}}\Big )^{c_4},
\label{eq:fit}
\end{equation}
where $a$ is the expansion scale factor, $d$ is the distance from the \ac{MW}-mass halo center, $M$ is the threshold in instantaneous subhalo mass, $d_{0}$ is a unit normalization of $1 \kpc$, and $M_{0} = 10^{7} \Msun$ is our fiducial mass threshold.
Table~\ref{tab:fit-parameters} lists the best-fit parameters, $c_0$, $c_1$, $c_2$, and $c_3$, for each metric.
We generated each of these constants from a particular curve using the Levenberg–Marquardt algorithm: $c_{0}$ and $c_{3}$ from $M_\textrm{sub} > 10^{7}$ at $z = 0$ (orange curve in Figure~\ref{masspop}), and $c_{1}$ and $c_{2}$ from $M_\textrm{sub} > 10^{7} \Msun$ at $z = 1$ (brown curve in Figure~\ref{zpop}).
We also fit constant $c_4$ to subhalos with $M_{\textrm{sub}} > 10^{8} \Msun$ at $z = 0$ (green curve in Figure~\ref{masspop}). However, $c_4$ values for all metrics were both very close to one another and very close to previously determined values for the subhalo mass function \citep{Wetzel2009}. In order to reduce the number of constants in the function, we use the median value of our three population metrics, $c_4=-0.94$, consistent with longstanding expectations for the subhalo mass function.
As a metric of uncertainty in our fit, we also include the fractional uncertainty in population amplitude $c_0$, found by taking the product of $c_0$ with the average host-to-host scatter for our fiducial mass range $M_{\rm sub}> 10^{7\ \Msun}$ at $z=0$ (orange curves in Figure~\ref{masspop}).

We indicate the distance region used for each fit in Figure~\ref{masspop} and Figure~\ref{zpop} in color; curves are shown in grayscale outside of this region.
We did not use results for $M_\textrm{sub} > 10^{6} \Msun$ to fit any parameters, given possible limitations from numerical resolution (see Appendix~\ref{sec:app}).
However, the dotted lines in Figures~\ref{masspop} and \ref{zpop} show that our fits for this mass threshold are generally within the 68 per cent host-to-host scatter at $d < 50 \kpc$, reinforcing that any numerical underestimate is likely less than a factor of $\approx 2$.

\subsection{Orbital velocity distributions}
\label{sec:velocity}

\begin{figure*}
\centering
\begin{tabular}{c}
\includegraphics[width=0.94\linewidth]{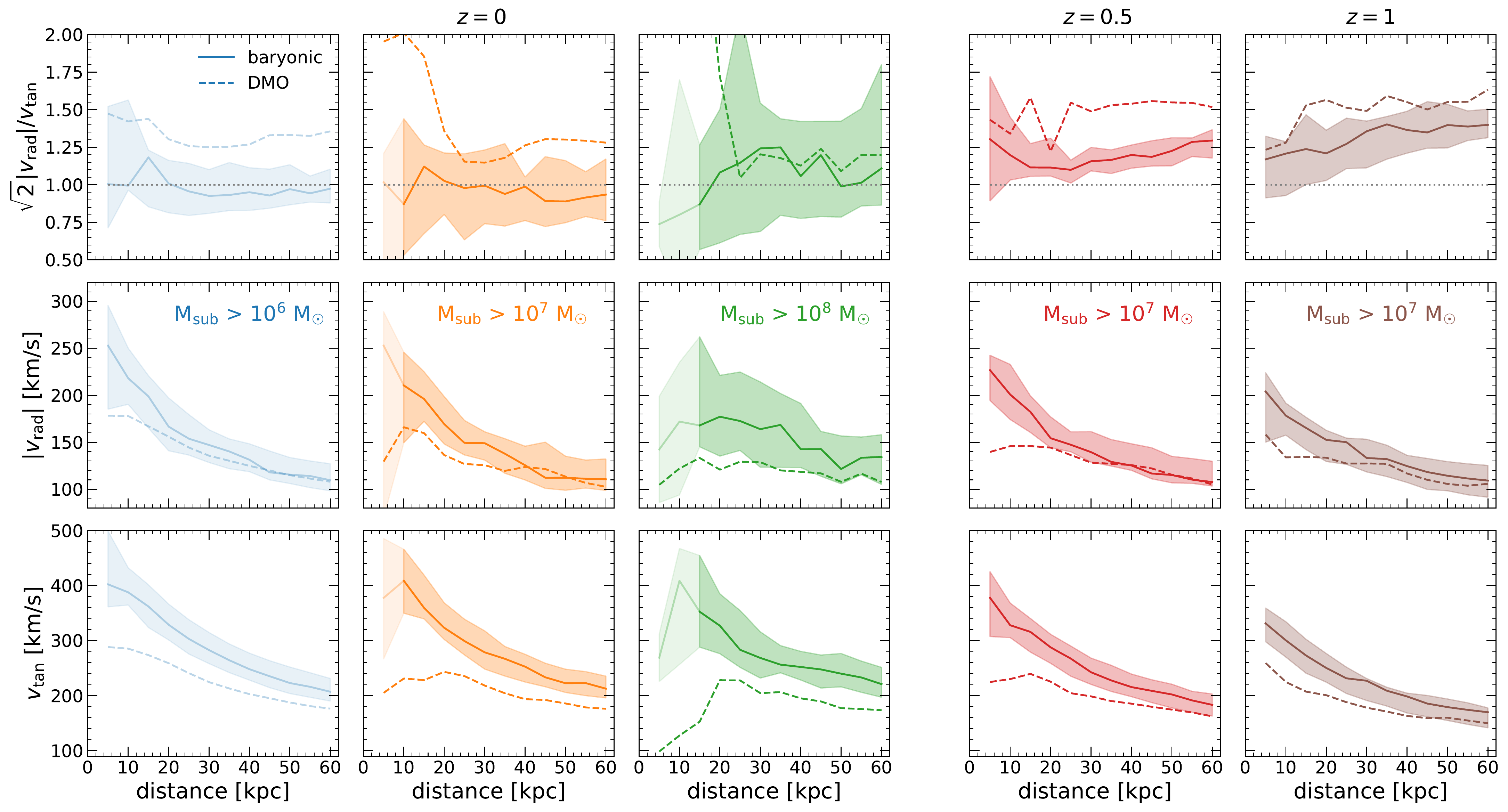}
\end{tabular}
\vspace{-1 mm}
\caption{
Orbital velocities of subhaloes versus distance, $d$, from the \ac{MW}-mass galaxy.
Solid lines show the mean, and shaded regions show the $68$ per cent distribution across the 11 host haloes, while dashed lines show dark-matter-only (DMO) simulations of the same haloes. Dashed lines for $z=0.5$ and $z=1$ show only \textit{Latte} hosts. Lighter shade shows bins where more than 1 halo had an average of 0 subhaloes.
\textbf{Left}: subhaloes at $z \approx 0$, for $M_\textrm{sub} > 10^{6}$, $10^{7}$, and $10^{8} \Msun$.
\textbf{Right}: subhaloes with $M_\textrm{sub} > 10^{7} \Msun$ at $z \approx 0$, 0.5, and 1.
\textbf{Top}: Orbital velocity isotropy, via the dimensionless ratio of the median absolute radial velocity to the median tangential velocity, normalized such that isotropic orbits have a value of 1. While subhaloes in DMO simulations have radially biased orbits at all redshifts, subhaloes in baryonic simulations orbit in a nearly statistically isotropic distribution at $z \approx 0$. At higher redshifts, subhalo orbits in baryonic simulations were increasingly radially biased.
\textbf{Middle}: Median absolute radial velocity, $|v_{\rm rad}|$. The deepening of the gravitational potential from the \ac{MW}-mass galaxy in the baryonic simulations increases $v_{\rm rad}$ at small $d$ relative to DMO, but the two are nearly identical at $d \gtrsim 40 \kpc$.
\textbf{Bottom}: Tangential velocity, $v_{\rm tan}$, is higher in baryonic simulations than in DMO, and this enhancement persists at all $d$. In addition to the deepening of the gravitational potential, as above, subhaloes with small $v_{\rm tan}$ are more likely to get tidally stripped by the host galaxy and fall below the mass threshold, which further enhances $v_{\rm tan}$ of the surviving population.
Thus, subhalo orbits are statistically isotropic at $z \lesssim 0.5$ ($t^\textrm{lb} \lesssim 6 \Gyr$).
}
\label{vel}
\end{figure*}

An essential component for modeling subhalo-stream interactions is the direction of the subhalo orbit relative to the stream \citep{Yoon_2011, Banik:2018pjp}. Figure~\ref{vel} shows the subhalo orbital velocity components across varying masses and redshifts.
The first row shows a metric of the orbital isotropy of the subhalo population, via the ratio of the average absolute radial velocity, $|v_\textrm{rad}|$, to the average tangential velocity, $v_\textrm{tan}$, normalized so that unity represents statistically isotropic orbits. The left three columns show results at $z \approx 0$ for our three thresholds in instantaneous mass.
The right two columns show subhaloes of $M_\textrm{sub} > 10^{7} \Msun$ at $z \approx 0.5$ and $z\approx 1$, as in Figure~\ref{zpop}.

At $z \approx 0$, subhaloes in baryonic simulations are consistent with isotropic orbits, in contrast to DMO simulations, in which subhalo orbits are radially biased.
Our results suggest that one can approximate a statistically isotropic velocity distribution when modeling and interpreting possible orbits for subhaloes at a given $d$ at $z \approx 0$.
However, this was not always true: at earlier cosmic times, subhaloes in baryonic simulations were somewhat more radially biased, by up to $1.3 \times$ at $z = 0.5$ and up to $1.4 \times$ at $z = 1$, with larger radial bias at larger $d$.
The DMO simulations also had higher radial bias at earlier times.
Most likely, the higher radial bias at earlier cosmic times in both baryonic and DMO simulations arises because subhaloes necessarily fell in more recently, reflecting their initial infall orbits more directly \citep[for example][]{Wetzel2011}. Subhaloes that are on highly radial orbits also pass closer to host center and thus strip/merge more quickly.
That said, the reason why the additional gravitational effects of the \ac{MW}-mass galaxy in the baryonic simulations should lead to a surviving subhalo population with nearly isotropic orbits at $z \approx 0$ is not obvious; we defer a more in-depth investigation to future work.

To provide deeper insight into the orbital velocity isotropy, the bottom two rows of Figure~\ref{vel} show the individual velocity components, $|v_\textrm{rad}|$ and $v_\textrm{tan}$. Beyond $\approx 40 \kpc$, $|v_{\rm rad}|$ is similar in both baryonic and DMO simulations.
In both, $|v_{\rm rad}|$ increases at smaller $d$, where the gravitational potential is deeper. However, $|v_{\rm rad}|$ is larger at small $d$ in the baryonic simulations, because the formation of a \ac{MW}-mass galaxy deepens the potential.

In the bottom row, $v_{\rm tan}$ is higher in baryonic simulations at all $d$, though again the enhancement is most significant at small $d$. In addition to the host galaxy deepening the potential, it also provides additional gravitational tidal stripping for subhaloes that orbit close to it, which have small orbital angular momentum, as \cite{GK2017} showed. This in turn biases the resultant subhalo population at a given $d$ to have higher $v_{\rm tan}$.
Thus, the stronger enhancement of $v_{\rm tan}$ leads to the change from radially biased orbits in DMO to statistically isotropic orbits in baryonic simulations for surviving subhaloes above a given mass threshold.

At earlier times, both $|v_\textrm{rad}|$ and $v_\textrm{tan}$ in baryonic simulations were more similar to those in DMO simulations than they are at $z \approx 0$, demonstrating how the tidal effects of the host galaxy affected the subhalo population over time.
The host galaxy stellar mass increased significantly over this time interval: relative to its stellar mass at $z = 0$, it typically was only half as large at $z = 0.5$ and only about a quarter as large at $z = 1$ \citep{Santistevan2020, Bellardini2022}.

\subsection{Subhalo enhancement during LMC passage}
\label{sec:LMC}

\begin{figure}
\centering
\begin{tabular}{c}
\includegraphics[width = 0.94 \linewidth]{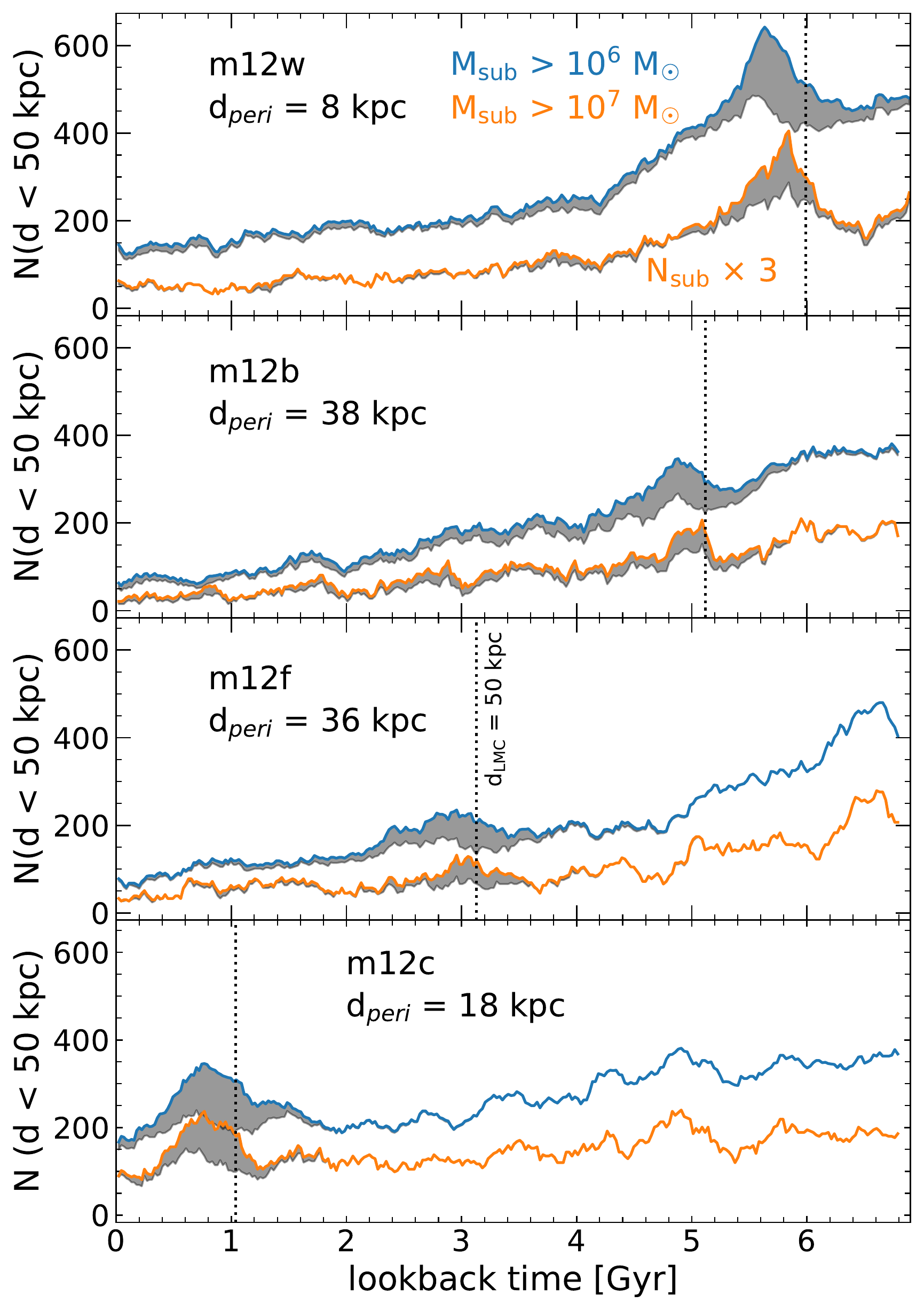}
\end{tabular}
\vspace{-1 mm}
\caption{
Number of subhaloes within $d < 50 \kpc$ of the \ac{MW}-mass galaxy versus cosmic time, for the 4 simulations that have an LMC satellite analog.
We show subhaloes with instantaneous mass $> 10^{6}$ and $> 10^{7} \Msun$, with the latter multiplied by 3 for clarity. Grey shaded regions show subhaloes that were satellites of the LMC analog any time prior to infall. Vertical dotted lines shows when the LMC analog first orbited within 50 kpc of the \ac{MW}-mass galaxy, which is the current distance of the LMC from the \ac{MW}. All 4 cases show significant enhancement in subhaloes for $\approx 1-2 \Gyr$ after first infall, after which orbital phase mixing leaves no coherent enhancement during subsequent pericentric passages.
Figure~\ref{lmc_ratio} and Table~\ref{tab:LMC-enhancement} quantify the enhancement of subhaloes during LMC passage.
}
\label{lmc}
\end{figure}

\begin{table*}
\centering
\caption{
Enhancement of the number and orbital radial flux of subhaloes within $50 \kpc$ of the \ac{MW}-mass host during 4 LMC satellite analog events (see Table~\ref{tab:LMC-table}). We measure the subhalo population within $\pm 50 \Myr$ of when each LMC analog first reached $d = 50 \kpc$, and we show the mean and standard deviation of the ratio (as defined below) across these 4 LMC analog events.
\textbf{Top rows}: ratio of each \ac{MW}-mass halo at the time the LMC analog reached $d = 50 \kpc$ to the average for the same \ac{MW}-mass halo $100 - 500 \Myr$ prior, before LMC infall.
Subhalo counts and fluxes show a consistent enhancement ($1.1 - 2.3 \times$), so the infall of the LMC analog significantly boosted the host halo's subhalo population.
\textbf{Bottom rows}: ratio of each \ac{MW}-mass halo with an LMC analog to the average of all 11 \ac{MW}-mass haloes at the same redshift.
Subhalo counts and fluxes show similar enhancements ($1.4 - 2.1 \times$).
\textit{Thus, the \ac{MW} today likely has a significantly enhanced ($\approx 2 \times$) population of subhaloes relative to similar-mass host haloes today and relative to its own population several 100 Myr ago.}
}
\begin{tabular}{llllll}
\hline
& subhalo mass threshold [$\Msun$] & number enhancement & flux enhancement\\
\hline 
relative to same \ac{MW}-mass halo, & $>10^6$ & 1.12 $\pm$ 0.05 & 1.11 $\pm$ 0.12 \\
$100 - 500$ Myr prior & $>10^7$ & 1.40 $\pm$ 0.08 & 1.37 $\pm$ 0.21 \\
& $>10^8$ & 1.42 $\pm$ 0.32 & 1.38 $\pm$ 0.59 \\
\hline
relative to all \ac{MW}-mass haloes & $>10^6$ & 1.41 $\pm$ 0.49 & 1.64 $\pm$ 0.53 \\
at same redshift & $>10^7$ & 1.83 $\pm$ 0.72 & 2.00 $\pm$ 0.78 \\
& $>10^8$ & 2.15 $\pm$ 1.15 & 1.56 $\pm$ 1.16 \\
\hline
\end{tabular} 
\label{tab:LMC-enhancement}
\end{table*}

\begin{figure}
\centering
\begin{tabular}{c}
\includegraphics[width = 0.94 \linewidth]{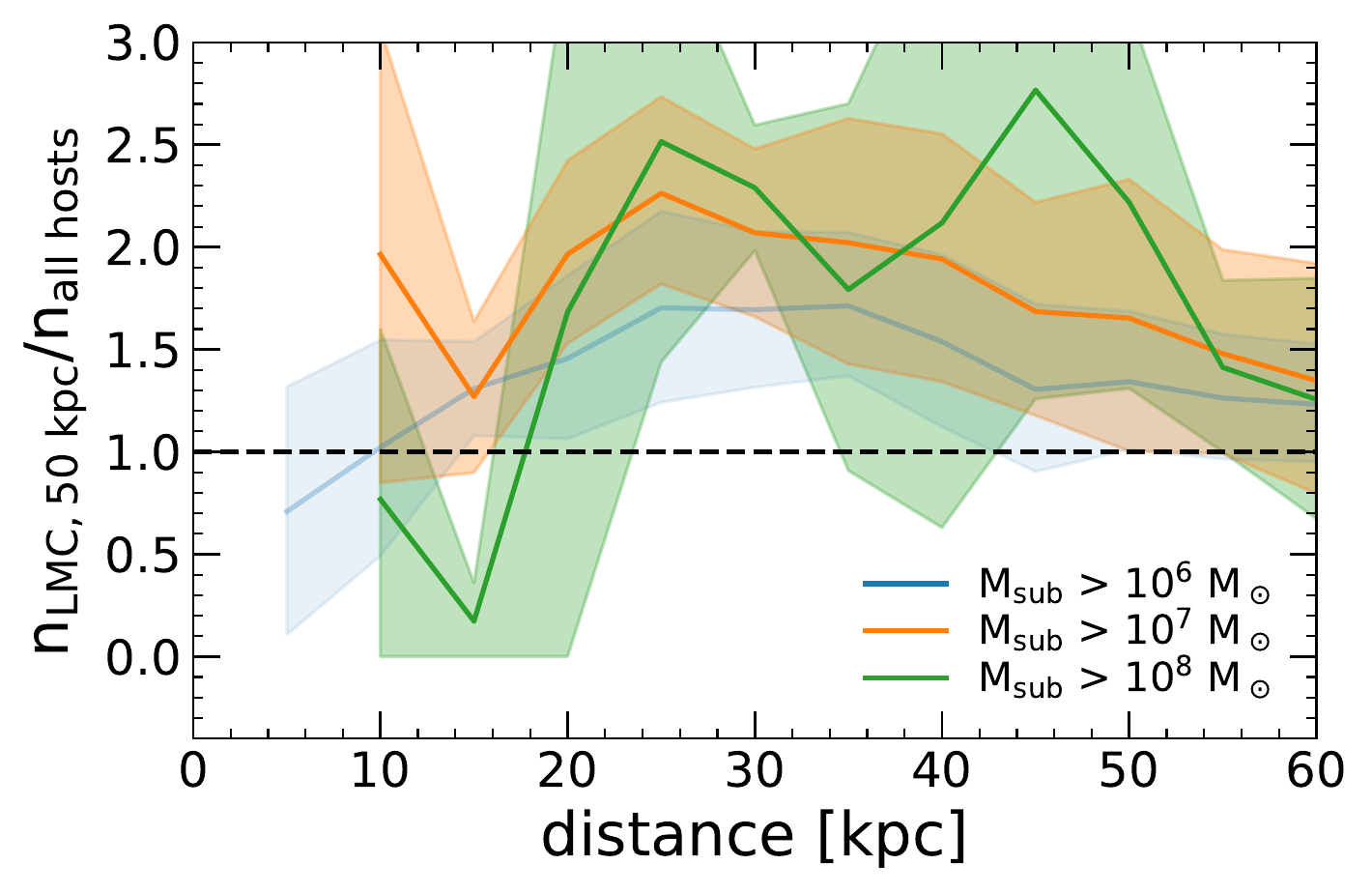}
\end{tabular}
\vspace{-3 mm}
\caption{
The enhancement of subhaloes around \ac{MW}-mass galaxies with an LMC satellite analog.
We compute the ratio of the average number density, $n(d)$, of subhaloes around each of the 4 hosts with an LMC satellite analog (m12w, m12c, m12f, and m12b), time-averaged over $\pm 50$ Myr when the LMC analog fist crossed within $d = 50 \kpc$, relative to the average across all 11 \ac{MW}-mass haloes at the same redshift. Shaded regions show the standard deviation across the 4 hosts.
\textit{\ac{MW}-mass haloes with an LMC satellite analog show a strong enhancement, typically $1.4 - 2.1 \times$, with only a weak decline with distance.}
}
\label{lmc_ratio}
\end{figure}

To predict the current subhalo population around the \ac{MW}, we examine the potential impact of the \ac{LMC}, a massive satellite galaxy ($M_\textrm{DM} \sim 10^{11} \Msun$) that recently passed its pericenter of 50 kpc \cite{Kallivayalil2013}. We focus on the 4 simulations with LMC satellite analogs in Section~\ref{sec:lmcselection}: m12w, m12b, m12f, and m12c.

Figure~\ref{lmc} shows the subhalo counts over time in each simulation, quantified as the cumulative number of subhaloes within 50 kpc of the \ac{MW}-mass host, for $M_{\rm sub} > 10^{6} \Msun$ and $M_{\rm sub} > 10^{7} \Msun$. Counts for both mass thresholds visibly increased during the $\sim 50 \Myr$ after the LMC analog first reached $d = 50 \kpc$, which we indicate with a dotted line. We do not show subhaloes $> 10^{8} \Msun$ because of their low counts ($\lesssim 10$ subhaloes at any given time) and therefore significant Poisson scatter, but they show similar increases in all 4 simulations.

The grey shaded region indicates the number of subhaloes that \rockstar\ identifies as having been a satellite of the LMC analog halo any time before becoming a satellite of the \ac{MW}-mass halo, demonstrating that this enhancement is primarily (though not entirely) from subhaloes that were satellites of the LMC analog. This period of subhalo enrichment lasts for only $\lesssim 0.5 \Gyr$ after the LMC analog's first pericentric passage, consistent with previous works that have shown that satellites of LMC-mass satellite galaxies phase mix on this timescale \citep[for example][]{Deason2015}.
While some of these additional subhaloes persist indefinitely, the subsequent phase mixing of their orbits leads to no strong temporal enhancement.
While these LMC analogs have smaller pericenters about their \ac{MW}-mass host ($8 - 38 \kpc$) than the $d_{\rm peri} \approx 50 \kpc$ of the LMC, all 4 show significant enhancement already when the LMC analog first crosses within $d = 50 \kpc$ (vertical dotted line).
The latest \ac{LMC} analog to reach $d = 50 \kpc$ is in m12c at $z = 0.07$ ($t^{\rm lb} = 0.95 \Gyr$), making it temporally the most similar to the \ac{LMC}; subhalo counts in this host are $2 - 4 \times$ higher than in the other \ac{MW}-mass hosts at the same redshift.

Table~\ref{tab:LMC-enhancement} quantifies the enhancement in the cumulative number and the orbital radial fluxes of subhaloes in our hosts with an \ac{LMC} satellite analog via two ratios. The first row compares subhalo counts in each of the four hosts with an \ac{LMC} analog at $t_{\rm 50 \kpc}$, the time at which the \ac{LMC} analog first reached $d = 50 \kpc$, to the value in the same host $100 - 500 \Myr$ earlier. The second row compares subhalo counts in each host with an \ac{LMC} analog within $\pm 50 \Myr$ of $t_{\rm 50 \kpc}$ to the average counts in all 11 \ac{MW}-mass hosts at the same time. We measure cumulative number as in Figure~\ref{masspop}: the total number of subhaloes enclosed within a given distance, in this case $d=50\ \rm kpc$. We measure the flux similarly to Figure~\ref{masspop}, now taking the average value of all flux rate data points within the range $d=7.5 - 50\ \textrm{kpc}$, representing the flux enhancement over the full range of distances relevant to stellar stream detections.
Both the subhalo counts and fluxes increase $\approx 1.4 - 2.1 \times$, with broadly consistent enhancement across different mass thresholds.
Within $R_{\rm 200m}$, the enhancement in absolute number is much higher ($\approx 100$ for $M_{\rm sub} > 10^{7} \Msun$) than for $d < 50$ kpc ($\approx 30$), which means that only a fraction of the subhaloes that accreted with the LMC analog contribute to our results at $d < 50$ kpc.
However, the \textit{fractional} enhancement inside $R_{\rm 200m}$ (relative to other hosts at the same time) is weaker than inside $d < 50$ kpc, being $\approx 1.1 \times$ at all masses, because of the much larger number of preexisting subhaloes within $R_{\rm 200m}$ than within $d < 50$ kpc.

Figure~\ref{lmc_ratio} shows the relative enhancement in subhalo number density, $n(d)$, as a function of $d$, within our hosts with an LMC analog within $\pm 50$ Myr of $t_{\rm 50 \kpc}$, compared to all other hosts at the same redshift (as in Table~\ref{tab:LMC-enhancement}, row 2). We find a typical enhancement of $\sim 1.5 - 2 \times$ at all distances $> 10 \kpc$, with relatively weak dependence on both distance and subhalo mass.

Given that the \ac{LMC} is just past its first pericentric passage, our results imply that the \ac{MW} currently is experiencing a significant boost, typically $1.4 - 2.1 \times$, in its population of subhaloes at distances $\lesssim 50 \kpc$, both relative to itself a few hundred Myr earlier, and relative to other \ac{MW}-mass host haloes without an LMC analog at $z = 0$.
Thus, in making predictions of subhalo counts around the \ac{MW} today, one should multiply our host-averaged fits in Equation~\ref{eq:fit} by $\approx 2 \times$ (Table~\ref{tab:fit-parameters}).

\subsection{Predictions for interaction rates with stellar streams}
\label{sec:predictions}

We conclude by synthesizing our results to make approximate estimates for the interaction rates of subhaloes with stellar streams around the \ac{MW}. To reduce uncertainties from the details of subhalo-stream interaction dynamics, we consider here only near-direct collisions between subhaloes and streams; thus, these values represent conservative estimates of interaction rates.

As case studies, we use the fiducial streams, GD-1 ($d_{\rm peri} = 13 \kpc$, $d_{\rm apo} = 27 \kpc$, length $l = 15 \kpc$) and Pal 5 ($d_{\rm peri} = 8 \kpc$, $d_{\rm apo} = 19 \kpc$, $l = 10 \kpc$), approximating each as a thin cylinder.
We use relevant impact parameters ($b$) for potentially observable subhalo interactions with streams, for each of our three mass thresholds, from \cite{Yoon_2011}: $b < 0.58 \kpc$ for $M_\textrm{sub} > 10^{6} \Msun$, $b < 1.6 \kpc$ for $M_\textrm{sub} > 10^{7} \Msun$, and $b < 4.5 \kpc$ for $M_\textrm{sub} > 10^{8} \Msun$. These impact parameters come from the tidal radii of subhaloes in each mass bin, thus representing direct subhalo-stream impacts.
We then compute the average subhalo flux from our results (Figure~\ref{masspop}, bottom panel) across galactocentric distances corresponding to the orbital ranges of GD-1 ($13-27\ \rm kpc$) and Pal 5 ($8-19\ \rm kpc$). We reiterate that, because subhalo orbits are largely isotropic at $z\approx 0$, this radial flux rate can be used as the flux rate in any direction.

We use interaction rates at $z < 0.15$ ($t^{\rm lb} = 0.15 \Gyr$), and apply an enhancement from the LMC corresponding to the orbital range of each stream, as in Table~\ref{tab:LMC-enhancement}.
Under these conditions,
\textit{
for GD-1 we estimate $\approx 4 - 5$ interactions per Gyr with subhaloes $> 10^{6} \Msun$, $\approx 1 - 2$ per Gyr with $> 10^{7} \Msun$, and $\approx 0 - 1$ per Gyr with $> 10^{8} \Msun$;
for Pal 5, we estimate $\approx 2 - 3$ interactions per Gyr with subhaloes $ > 10^{6} \Msun$, $\approx 0 - 1$ per Gyr with $\it > 10^{7} \Msun$, and $\approx 0 - 1$ per Gyr with $\it > 10^{8} \Msun$.
}
If observable features in streams, such as gaps, persist for many Gyrs, then the evolution across cosmic time is important to incorporate.
In this case, we can estimate the `effective' rate by averaging our fluxes across $z = 0 - 0.5$ ($t^\textrm{lb} = 0 - 5.1 \Gyr$), but now omitting the boost factor from the (recently accreted) LMC.
For each stream, this increases the interaction rate per Gyr by $\approx 2 \times$ for $M_\textrm{sub} > 10^{6}$ and $10^{7} \Msun$, and up to $4 \times$ for $M_\textrm{sub} > 10^{8} \Msun$. Including distant and indirect subhalo-stream encounters, in addition to direct collisions, would further increase these predicted rates.

While only estimates, these encounter rates offer context for our results.
Even with the additional tidal effects of the \ac{MW}-mass galaxy in baryonic simulations, which significantly reduces the population of subhaloes at these distances relative to DMO simulations (down to $10\%$ or fewer from their DMO equivalents), \textit{our results imply that, within CDM, interaction rates between stellar streams and dark subhaloes are still sufficient to leave detectable, tidally-induced features.}

\section{Summary and discussion}
\label{sec:disc}

\subsection{Summary of Results}

Using 11 \ac{MW}-mass host galaxies from the FIRE-2 suite of cosmological simulations, we presented predictions for the counts and orbital distributions of low-mass subhaloes at $d \lesssim 50 \kpc$ around the \ac{MW} and \ac{MW}-mass galaxies.
Our primary goal is to inform studies that model potentially observable interactions between such subhaloes and stellar streams.
We explored the dependence on subhalo mass, distance, redshift, and the presence of an \ac{LMC} satellite analog, and we provided analytic fits to these dependencies.

Our primary results are:
\begin{enumerate}[leftmargin=*]
\item The incorporation of baryonic physics significantly reduces subhalo counts compared with DMO simulations, primarily because of additional tidal force from the \ac{MW}-mass galaxy potential.
At $z \approx 0$, DMO simulations overpredict subhalo counts by $\approx 4 - 5$ times at $d \approx 20$ kpc.
These differences were less pronounced at earlier cosmic times, with DMO simulations overpredicting counts by $\approx 1.5$ times at $z = 0.5$.
\item \textit{We predict that $> 20$ ($> 4$) subhaloes with instantaneous mass $> 10^{6} \Msun$ ($> 10^{7} \Msun$) exist within the distances of streams like GD-1 and Pal 5 ($d \lesssim 30 \kpc$), and at least 1 subhalo $>10^{8} \Msun$ resides within $d < 50 \kpc$.}
Thus, despite the strong depletion of subhaloes in baryonic simulations relative to DMO, significant numbers of subhaloes survive at the distances of observed stellar streams.
This is unlike \cite{GK2017}, who found no surviving subhaloes within $\approx 15$ kpc, but they only examined two of these FIRE-2 simulations (m12i, m12f) at a single snapshot at $z = 0$.
\item At $z \approx 0$, subhalo number density and orbital flux are nearly constant with distance, out to at least $d \approx 60 \kpc$.
\item Subhalo counts decreased significantly over cosmic time, from both the declining rate of infall of new subhaloes and the increasingly strong tidal field of the host galaxy. At $z = 1$, the \ac{MW}-mass hosts had $\approx 10$ times more subhaloes at a given distance.
This decline over time is stronger at smaller distances and at higher subhalo masses.
\item \textit{Subhaloes orbit with statistically isotropic velocities at $z \approx 0$} but they were increasingly radially biased at earlier times.
This is unlike DMO simulations, in which subhalo orbits are always radially biased. 
That subhalo velocities are isotropic at $z\approx 0$ implies that our subhalo radial flux values can also be applied to subhalo flux in any desired direction.
\item \textit{The initial infall of an LMC satellite analog boosts the number of subhaloes within 50 kpc of the \ac{MW}-mass host by $1.4 - 2.1$ times, relative to the same host a few hundred Myr earlier or relative to similar-mass host haloes at the same time.}
Thus, predictions and models for subhalo-stream interaction rates over the last few 100 Myr should take into account this enhancement from the \ac{LMC}.
\end{enumerate}

\subsection{Discussion}

As expected, we find similar overall results as \cite{GK2017}, who examined two of the same FIRE-2 haloes as we do (m12i, m12f). However, we emphasize key differences and extensions of our work compared with theirs. First, we include more host haloes (11) for better statistics. Second, and equally importantly, we time-averaged our results across multiple snapshots. Given our snapshot time spacing of $20 - 25 \Myr$, this provides a much more statistically representative picture of subhaloes at small distances, where their velocities are highest and where they spend the least time in their orbit. Furthermore, we interpolate subhalo distances between snapshots (see Section~\ref{sh_select}) to avoid missing orbits at particularly small distances. Unlike \cite{GK2017}, who found no surviving subhaloes within $\approx 15$ kpc of m12i and m12f at $z = 0$, we find that subhaloes survive, if briefly, at all $d \gtrsim 5 - 10 \kpc$ at all mass thresholds.

We next discuss caveats to our results. While we selected these host galaxies/haloes for their similarity to the \ac{MW}, they are not exact analogs. Each one has a different formation and merger history, resulting in significant host-to-host variation \citep{Santistevan2020}. In general, we averaged our results across these 11 hosts and included the host-to-host scatter, to present cosmologically representative results for subhaloes around \ac{MW}-mass galaxies.
This is statistically likely to encompass the population around the \ac{MW}, but there is no guarantee of it. We compared our results for our isolated haloes to our Local Group analogs and found negligible differences, consistent with the comparisons among the ELVIS DMO simulations in \cite{Wetzel2015}, indicating that such environmental selection is not a significant factor affecting these low-mass subhaloes at distances $\lesssim 50 \kpc$.

More critically, our results demonstrate that the presence of an \ac{LMC} satellite analog boosts low-mass subhalo counts by $1.4-2.1 \times$ at distances $\lesssim 50 \kpc$ (Table~\ref{tab:LMC-enhancement}), indicating that this is one of, and likely the, most important factor in predicting subhalo populations around the \ac{MW} today and over the last few 100 Myr. 
All of our LMC analogs have smaller pericenters than the LMC, ranging from $8 - 38 \kpc$. We mitigated this by measuring subhaloes when the LMC analog first cross within the LMC's current pericentric distance of $\approx 50 \kpc$.
Arora et al. (in preparation) examine subhalo populations in selected hosts of the FIRE-2 simulations at different angular locations around the galactic center, including the spatial relation to LMC analogs, as well as specific subhalo-stream encounter rates in the presence of a massive satellite.

We examined results only for one dark-matter model, CDM, but there are many other possible candidates. Extensions of our work would include examining FIRE simulations with alternative dark-matter models \citep[such as][]{Robles_2017, Shen2022}.

As with any simulation, our results are susceptible to resolution effects.
We reiterate that, motivated by quantifying the strength of gravitational interactions with stellar streams, we selected subhaloes above a given \textit{instantaneous} threshold in mass, so numerical convergence requires that our simulations accurately model the amount of mass stripping in subhaloes down to our given instantaneous mass threshold(s).
Thus, our results are not sensitive to modeling any mass stripping (physical or numerical) that occurs below this threshold, or to the more challenging question of modeling/defining subhalo `disruption', which occurs below these mass thresholds.

In Appendix~\ref{sec:app}, we quantify resolution convergence by comparing our fiducial subhalo counts to those in both lower-resolution and higher-resolution versions of the same haloes.
To summarize, our tests indicate that subhalo counts at instantaneous $M_\textrm{sub} > 10^{7} \Msun$ are reasonably well converged, but at $M_\textrm{sub} > 10^{6} \Msun$ our simulations underpredict the counts by up to $\approx 1.5 - 2 \times$ (which we have indicated throughout), so our results there are lower limits, which means that the actual interaction rates with streams would be even higher.
And again, in fitting our results, we did not include any values at $M_\textrm{sub} > 10^{6} \Msun$, so our fit values there are extrapolations from our higher masses.

We also discuss the numerical convergence of our subhaloes in the context of the criteria that \cite{vdbosch-2018} provided from idealized simulations of individual subhaloes orbiting in fixed host halo potential without a central disk.
They consider a subhalo to be sufficiently resolved based on its bound mass fraction, $f_{\rm bound}$, the ratio of its instantaneous mass to its peak mass (typically just before accretion), with a minimum $f_{\rm bound}$ determined by the subhalo's scale radius, $r_{\rm s,0}$, and the number of DM particles it had at its peak mass, $N_{\rm peak}$.
They define $f_{\rm bound}^{\rm min,1} = C \left(\epsilon / r_{\rm s,0} \right)^{2}$ and $f_{\rm bound}^{\rm min,2} = 0.32 \left( N_{\rm peak} / 1000 \right)^{-0.8}$, where $C$ is a constant that depends on the subhalo's concentration parameter (we use $C \approx 10$, based on their Section 6.4), $\epsilon$ is the Plummer force softening of the simulation, which is $40 \pc$ for all of our simulations, and $N_{\rm peak}$ is the peak number of constituent DM particles a subhalo experienced, typically prior to accretion. \cite{vdbosch-2018} consider a subhalo converged if it satisfies both criteria, that is, $f_{\rm bound}^{\rm min} = \textrm{MAX}(f_{\rm bound}^{\rm min,1},f_{\rm bound}^{\rm min,2})$.

For reference, we note the median values for these relevant quantities for each of our subhalo samples at $z = 0$. For $M_{\rm sub} > 10^{6} \Msun$ (1,436 subhaloes), the median $M = 2.4 \times 10^{6} \Msun$, $M_{\rm peak} = 1.3 \times 10^{7} \Msun$, $r_{\rm s,0} = 0.15 \kpc$, and $N_{\rm peak} = 517$. For $M_{\rm sub} > 10^{7} \Msun$ (209 subhaloes), the median $M = 2.9 \times 10^{7} \Msun$, $M_{\rm peak} = 8.5 \times 10^{7} \Msun$, $r_{\rm s,0} = 0.18 \kpc$, and $N_{\rm peak} = 3419$. For $M_{\rm sub} > 10^{8} \Msun$ (13 subhaloes), the median $M = 2.3 \times 10^{8} \Msun$, $M_{\rm peak} = 8.2 \times 10^{8} \Msun$, $r_{\rm s,0} = 1.5 \kpc$, and $N_{\rm peak} = 3 9,461$. For all mass thresholds, the median $f_{\rm bound} \approx 0.26$, that is, all samples have experienced the same typical fraction of mass stripping since $M_{\rm peak}$.

Applying the convergence criterion from \cite{vdbosch-2018} to our samples, for $M_{\rm sub} > 10^{6}$, $> 10^{7}$, and $> 10^{8} \Msun$, the fraction of subhaloes that meet the criterion for mass resolution, $f_{\rm bound}^{\rm min,2}$, is $17$ per cent, $92$ per cent, and $100$ per cent, respectively.
The criterion for spatial resolution, $f_{\rm bound}^{\rm min,1}$, is more stringent.
Enforcing both $f_{\rm bound}^{\rm min,1}$ and $f_{\rm bound}^{\rm min,2}$ brings these fractions down to $6$ per cent, $39$ per cent, and $69$ per cent. Nearly all subhaloes at $M_{\rm sub} > 10^{7}$ and $10^{8} \Msun$ had $f_{\rm bound} > f_{\rm bound}^{\rm min,2}$ ($92$ per cent and $100$ per cent respectively), so $f_{\rm bound}^{\rm min,1}$ dominates this population's convergence fraction.
In agreement with our resolution tests, the convergence fraction for $M_{\rm sub} > 10^{6} \Msun$ is significantly lower.

However, the idealized simulations in \cite{vdbosch-2018} did not include a central disk potential, which significantly increases the physical tidal force and therefore mass stripping at $d \lesssim 50 \kpc$.
This may relax the criteria on $f_{\rm bound}^{\rm min}$; for example, in an extreme limit of a strong tidal field that induces (nearly) complete physical mass stripping at first pericenter, numerical considerations of resolving subhaloes above a given instantaneous mass threshold across many orbits become less significant.
\cite{Webb_2020} also explored the effects of resolution on simulated subhaloes, using re-simulations taken from the Via Lactea II simulation, and they also included a \ac{MW}-mass disk potential. They found that subhaloes with $M_{\rm sub} \sim 10^{7} \Msun$ at the resolution and force softening lengths of our FIRE-2 simulations can lose up to 60 per cent of their mass over the course of their lifetimes (across up to $\sim 5$ Gyr) relative to their counterparts at higher resolution, while subhaloes at $M_{\rm sub} \sim 10^{6} \Msun$ dissipate entirely.
Though, their static host galaxy potential had higher mass at earlier times than our cosmological simulations: over the last 5 Gyr, the central galaxy in our simulations increased by typically $\approx 30$ per cent. Additionally, individual subhalos exhibit a wide range of infall times; since most mass loss occurs during infall, subhalos with later infall times are subject to artifical mass loss effects for a shorter period of time. \cite{Santistevan23} examined the infall times of luminous satellites in the same simulations, finding a 68th percentile range of $4 - 10$ Gyr at the low subhalos masses that we analyze here.
We reiterate that our convergence tests in Appendix~\ref{sec:app} provide the most direct numerical test of our cosmological setup.



Comparing with previous works, our results generally agree with those that modeled a \ac{MW}-mass galaxy potential.
Compared to \cite{DOnghia10}, who examined subhaloes in the Aquarius DMO simulation with an added galaxy disk potential, we find the same order-of-magnitude results for $M_\textrm{sub} > 10^{6} \Msun$ (their counts being $\approx 2.5 \times$ higher than ours within $d = 50 \kpc$, and approximately the same as ours within $d = 20 \kpc$), but lower counts for $M_\textrm{sub} > 10^{8} \Msun$ ($\approx 4 \times$ within $d = 50$, $\approx 20 \times$ within $d = 20 \kpc$).
Other works that compared subhalo populations in DMO simulations to those that also model a central galaxy potential found that DMO simulations overpredict subhaloes at $d \lesssim 50 \kpc$ by $\approx 1.5 \times$ \citep{DOnghia10}, $\approx 1.8 \times$ \citep{GK2017}, $\approx 3.3 \times$ \citep{Kelley_2019}, and $\approx 3 \times$ \citep{Nadler2021}.
By, comparison, we find $\approx 2 - 3 \times$, on average, with some dependence on subhalo mass.
Additionally, \cite{Webb_2020} found that, broadly speaking, DMO simulations overpredict the \textit{entire} subhalo population within a \ac{MW}-mass halo by a factor of $\approx 1.6$, in broad agreement with our simulations, which have a mean DMO excess of $1.56$ for subhalos with $M_{\rm sub} > 10^{7} \Msun$ at $z=0$.

This reinforces that the most important effect of baryons for low-mass subhaloes is simply the addition of the tidal field from the central galaxy, as \citet{GK2017} demonstrated by showing similar results for the FIRE-2 baryonic simulations compared with simply adding a central galaxy potential to DMO simulations of the same haloes.
This agreement supports the use of an embedded central galaxy potential in DMO simulations as a computationally inexpensive alternative to simulations with full baryonic physics.
Furthermore, if using existing DMO simulations \citep[for example, as in][]{Hargis2014, Griffen2016}, one can increase the accuracy of subhalo counts by reducing them using the distance-dependent correction fits from \cite{Samuel2020}, which agree with our results, or using the machine-learning approach to subhalo orbital histories, as in \cite{Nadler2018}.


\cite{Sawala2017} examined subhaloes of instantaneous mass $10^{6.5} - 10^{8.5} \Msun$ in the APOSTLE simulations of Local Group analogs (DM particle mass $\approx 10^{4} \Msun$, force softening $\approx 134$ pc); we find broadly similar subhalo counts for both $M_\textrm{sub} > 10^{7} \Msun$ (within $\approx 1.4 \times$ of our counts) and $> 10^{8} \Msun$ (within $\approx 1.5 \times$ of our counts) at $d = 50 \kpc$.
\cite{Sawala2017} also compared their results to DMO versions of the same simulations and found similar DMO overpredictions of $\approx 2 \times$ at $d = 50 \kpc$ for $M_\textrm{sub} > 10^{7} \Msun$, with more dramatic DMO overprediction than our results at smaller distances ($\approx 4 \times$ at $d = 20 \kpc$).
However, the typical central galaxy in APOSTLE has significantly lower stellar mass, with $M_{\rm star} \approx 1.8 \times 10^{10} \Msun$, compared to our typical $M_{\rm star} \approx 6 \times 10^{10} \Msun$, which is similar to the \ac{MW}.
We also note similar trends in subhalo tangential and radial velocities, although subhalo orbits are generally less isotropic at small distances.

\cite{Zhu_2016} compared a baryonic versus DMO version of the Aquarius simulation, finding that DMO overpredicts subhaloes by $\approx 3 \times$ at $M_{\rm sub} > 10^{7} \Msun$ and $\approx 4 - 5 \times$ at $M_{\rm sub}> 10^{8} \Msun$ within the host halo's radius.
The larger-volume, lower-resolution Illustris and EAGLE simulations also demonstrate similar general trends of subhalo depletion in baryonic relative to DMO versions at small distances \citep[for example][]{Chua2017, Despali2017}.

We also compare to previous work that used simulations to predict subhalo populations and subhalo-stream interaction rates.
Our estimates for interaction rates (Section~\ref{sec:predictions}) are similar to those of \cite{Yoon_2011}, who used a lower-resolution DMO halo, designed to be similar to the \ac{MW}'s, with an added stream, predicted the Pal 5 stream to have $\approx 20$ detectable subhalo-induced gaps. \cite{Banik:2018pjp} simulated the evolution of GD-1 near a \ac{MW} potential and estimated that the \ac{MW} hosts $\approx 0.4 \times$ the number of subhaloes in a comparable DMO simulation, generally consistent with our results (Figure~\ref{masspop_ratio}).

To conclude, we presented cosmological predictions for subhalo counts and orbital fluxes (Figures~\ref{masspop} and \ref{zpop}) as well as velocity distributions (Figure~\ref{vel}), and we provided fits to these results, to inform studies that seek to predict and interpret observable effects of subhalo gravitational interactions on stellar streams.

\section*{Acknowledgments}
\label{sec:acknowledgements}

We thank Adrian Price-Whelan for suggesting that we explore the effect of the LMC on predictions for subhalo populations.
We thank Isaiah Santistevan, Pratik Gandhi, and Nicolas Garavito for helpful comments and discussion.
We also thank the anonymous referee for providing helpful and insightful comments and feedback.
MB and AW received support from: NSF via CAREER award AST-2045928 and grant AST-2107772; NASA ATP grant 80NSSC20K0513; HST grants AR-15809, GO-15902, GO-16273 from STScI.
We completed this work in part at the Aspen Center for Physics, supported by NSF grant PHY-1607611.
We ran simulations using: XSEDE, supported by NSF grant ACI-1548562; Blue Waters, supported by the NSF; Frontera allocations AST21010 and AST20016, supported by the NSF and TACC; Pleiades, via the NASA HEC program through the NAS Division at Ames Research Center.
We used NumPy \citep{numpy}, SciPy \citep{scipy}, AstroPy \citep{astropy:2018}, and MatPlotLib \citep{matplotlib}, as well as the publicly available package HaloAnalysis (\cite{wetzel2020}, available at \url{https://bitbucket.org/awetzel/halo_analysis/}).

\section*{Data Availability}
\label{sec:Data}

The data in these figures and all of the Python code that we used to generate these figures are available at the following repository: \url{https://bitbucket.org/meganbarry/subhalos_2023/}.
FIRE-2 simulations are publicly available \citep{Wetzel2022} at \url{http://flathub.flatironinstitute.org/fire}.
Additional FIRE simulation data is available at \url{https://fire.northwestern.edu/data}.
A public version of the \textsc{Gizmo} code is available at \url{http://www.tapir.caltech.edu/~phopkins/Site/GIZMO.html}.




\bibliographystyle{mnras}
\bibliography{bibliography} 



\appendix
\section{Resolution Convergence}
\label{sec:app}

\begin{figure}
\centering
\begin{tabular}{c}
\includegraphics[width = 0.94 \linewidth]{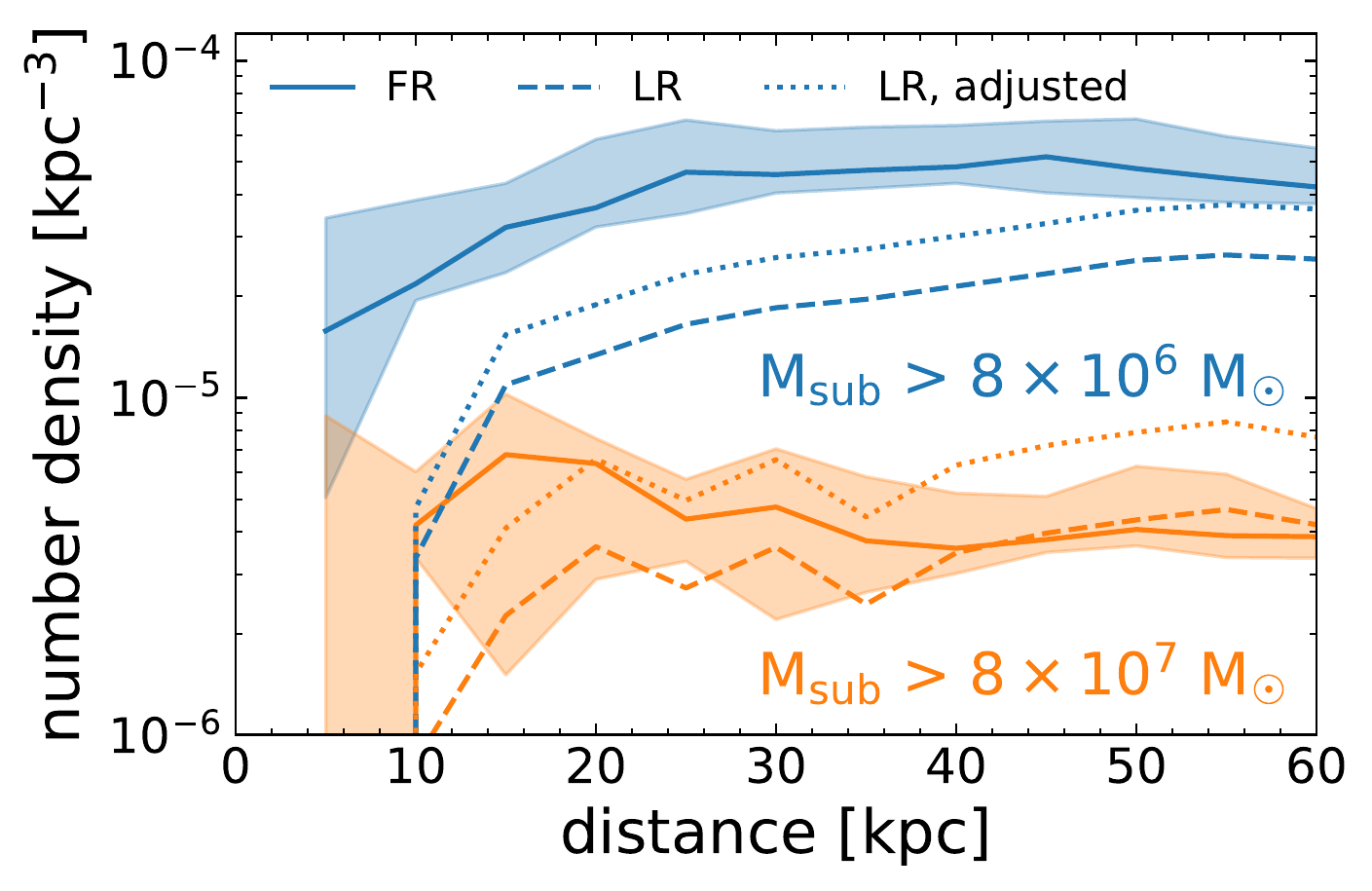}
\end{tabular}
\vspace{-3 mm}
\caption{
Resolution convergence test of subhalo number density, $n(d)$, versus distance from \ac{MW}-mass galaxy, $d$, for fiducial-resolution (FR) and lower-resolution (LR) simulations of the same 6 haloes in the \textit{Latte} suite, time-averaged over the same 92 snapshots as Figure~\ref{masspop}. LR simulations have $8 \times$ larger particle masses and $2 \times$ larger force softening. We show the average $n(d)$ and host-to-host scatter for $M_{\rm sub} > 8 \times 10^6 \Msun$ and $>8 \times 10^7 \Msun$, which correspond to the same number of particles in the LR simulations as for our fiducial mass thresholds of $>10^6 \Msun$ ($\approx 30$ particles) and $>10^7 \Msun$ ($\approx 300$ particles) in the FR simulations. Dotted lines show `adjusted' counts in the LR simulations to correct for them forming more massive host galaxies that induce stronger tidal stripping (see text). For subhaloes at $>8 \times 10^{7} \Msun$, $n(d)$ for LR and LR adjusted bracket and largely fall within the 68 per cent scatter of the FR simulations, especially at $d \approx 15 - 35 \kpc$.
By contrast, the LR simulations underpredict subhalo counts at $>8 \times 10^{6} \Msun$ by a factor of $\approx 2$. These results indicate that, in our FR simulations, subhalo counts at $M_{\rm sub} > 10^{7} \Msun$ are reasonably well converged but subhalo counts at $>10^{6} \Msun$ are lower limits.
}
\label{fig_append}
\end{figure}

\begin{figure}
\centering
\begin{tabular}{c}
\includegraphics[width = 0.94 \linewidth]{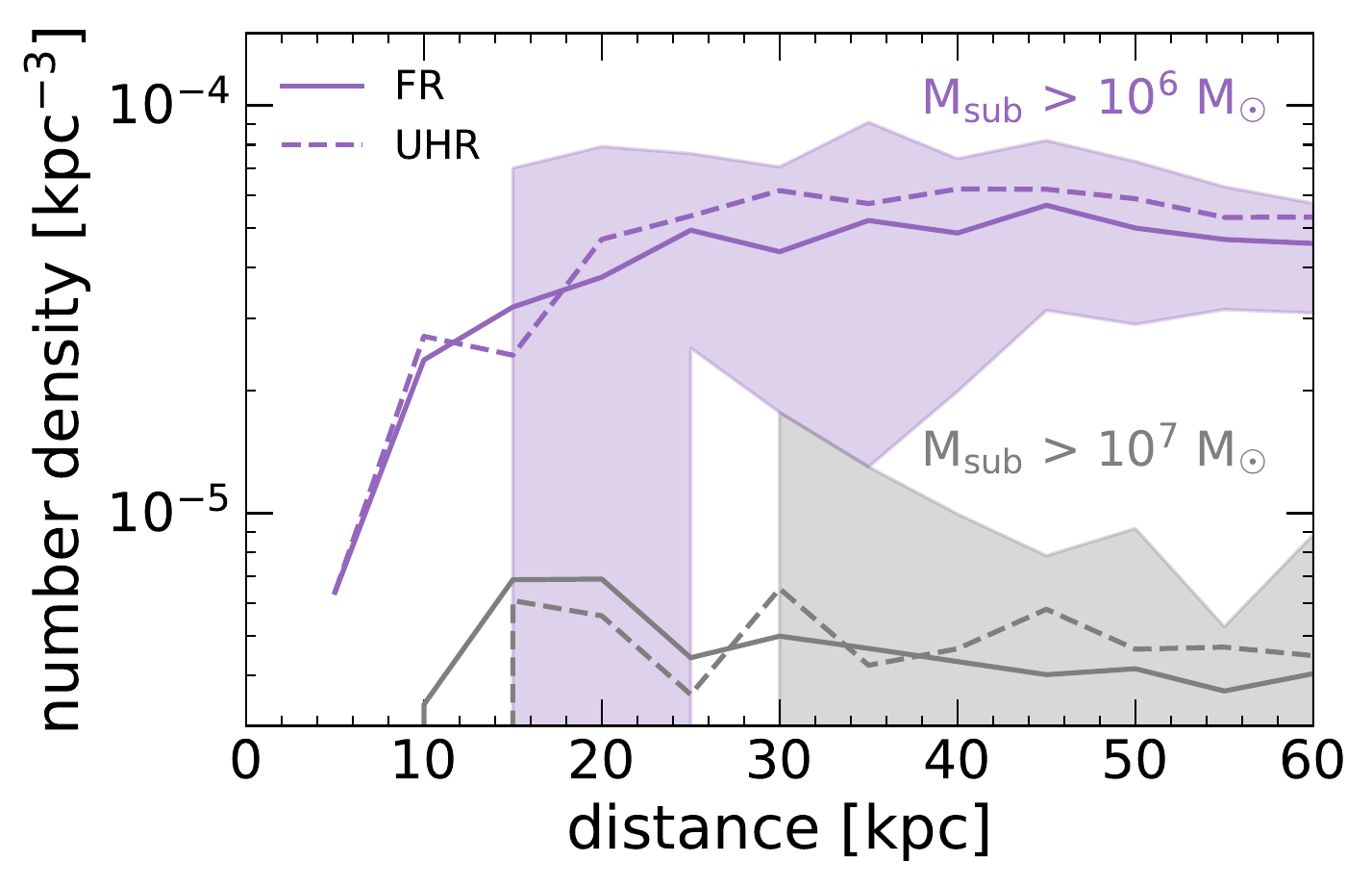}
\end{tabular}
\vspace{-3 mm}
\caption{
Resolution convergence test of subhalo number density, $n(d)$, versus distance from host halo center, $d$, for a single halo, m12i, time-averaged over the same 92 snapshots at $z \approx 0$ as Figure~\ref{masspop}.
We compare against re-simulations of m12i at ultra-high resolution (UHR), with $8 \times$ better mass resolution and $2 \times$ better force resolution than fiducial resolution (FR) (Wetzel et al., in preparation).
The shaded region shows the 68 per cent scatter across the 92 snapshots for our fiducial resolution.
Compared to Figure~\ref{fig_append}, we find better convergence between the two simulations as resolution increases, with differences between FR and UHR for $M_\textrm{sub} > 10^{6} \Msun$ being a factor of only $\approx 1.1 \times$.
}
\label{fig_append2}
\end{figure}

Here, we examine the resolution convergence of our counts for subhalos that we select above a given \textit{instantaneous} threshold in mass.
Thus, our tests are sensitive to how well the simulations model the correct amount of mass stripping that occurs down to a given instantaneous threshold in subhalo mass, but our results are not sensitive to mass stripping or `disruption' (physical or numerical) that occurs below that threshold.

We pursue two convergence tests.
First, we compare our fiducial-resolution (FR) simulations against a suite of lower-resolution (LR) re-simulations of the same haloes. We simulated each halo in the \textit{Latte} suite at 8 times lower mass resolution and 2 times larger force softening. Thus, subhaloes with $M_{\rm sub} > 8 \times 10^{6} \Msun$ and $M_{\rm sub} > 8 \times 10^{7} \Msun$ in the LR simulations have the same number of DM particles as those at $M_{\rm sub} > 10^{6} \Msun$ and $> 10^{7} \Msun$ in the FR simulations, so comparing counts at $M_{\rm sub} >8 \times 10^{6} \Msun$ and $> 8 \times 10^{7} \Msun$ provides a convergence test of our fiducial results at fixed number of DM particles (though the force softenings in LR simulations are also $2 \times$ larger). We do not test counts at $M_{\rm sub} > 8 \times 10^{8} \Msun$, because their small numbers lead to significant Poisson noise.

Figure~\ref{fig_append} shows subhalo number density, $n(d)$ (as defined in Section~\ref{sh_select}), as a function of distance from the host galaxy, $d$, time-averaged over the same 92 snapshots at $z \approx 0$ as in Figure~\ref{masspop}.

One complication to this comparison is that the LR simulations form host galaxies with higher ($\approx 1.7 \times$ on average) stellar mass \citep[see][]{Hopkins2018, Samuel2020}, resulting in an increased tidal force and stripping on subhaloes (in addition to resolution effects). Following \citet{Samuel2020}, we determine a correction factor for the subhalo count in the LR simulation, which we show as a dotted line (LR, adjusted). Specifically, we fit a power law to the relation between $N_{\rm sub}(< 50 \textrm{kpc})$ and host $M_\star$ at each threshold in subhalo mass.
This fit indicates that, at fixed resolution, a host galaxy with $1.7 \times$ larger stellar mass has 0.73 (0.58) times as many subhaloes at $> 8 \times 10^{6} \Msun$ ($> 8 \times 10^{7} \Msun$). Once we correct for the discrepancy in host galaxy mass, the subhalo counts at $M_\textrm{sub} > 8 \times 10^{7} \Msun$ agree reasonably well, within the host-to-host scatter, especially at $15 - 35 \kpc$, with a slight depletion at smaller $d$ and excess at larger $d$. Counts for subhaloes at $>8 \times 10^{6} \Msun$ are lower by a factor of $\approx 2$ in the LR simulations (adjusted for host galaxy mass) at all $d$, indicating that resolution effects are more important at this mass threshold.
Thus, as we emphasize throughout, we interpret predictions from our FR simulations at $M_\textrm{sub} > 10^{6} \Msun$ to be lower limits. See \cite{Samuel2020} for similar convergence tests comparing subhaloes at fixed peak mass (instead of instantaneous mass).

Second, for the single host halo m12i, we also compare our fiducial results against an ultra-high-resolution (UHR) version simulated to $z = 0$ (Wetzel et al., in prep.), for both baryonic and DMO simulations.
This UHR simulation has $8 \times$ smaller DM particle mass ($4400 \Msun$) and $2 \times$ smaller DM force softening ($20 \pc$). Figure~\ref{fig_append2} shows subhalo $n(d)$ versus $d$, comparing FR and UHR, for m12i at $z \approx 0$, time-averaged across the same 92 snapshots. In this case, the host galaxy forms the same stellar mass in both simulations, so we do apply any adjustment as in Figure~\ref{fig_append}.
Number density at both resolutions is now similar for both $>10^{6} \Msun$ and $>10^{7} \Msun$, to within the snapshot-to-snapshot scatter.

Taken together, Figures~\ref{fig_append} and \ref{fig_append2} imply that our FR simulations, which we presented throughout, have reasonably converged subhalo counts at $>10^{7} \Msun$, but they underpredict subhalo counts at $M_\textrm{sub} > 10^{6} \Msun$ by up to $\approx 1.5-2 \times$.

\bsp	
\label{lastpage}

\end{document}